\documentstyle[a4,epsf,psfig]{article}
\begin{document}

\title{Three-body halos. \\ 
V. Computations of continuum spectra for Borromean nuclei}

\author{A. Cobis, D. V. Fedorov and A. S. Jensen \\
Institute of Physics and Astronomy, \\
Aarhus University, DK-8000 Aarhus C, Denmark}

\maketitle

\begin{abstract}
We solve the coordinate space Faddeev equations in the continuum. We
employ hyperspherical coordinates and provide analytical expressions
allowing easy computation of the effective potentials at distances
much larger than the ranges of the interactions where only $s$-waves
in the different Jacobi coordinates couple.  Realistic computations
are carried out for the Borromean halo nuclei $^6$He (n+n+$\alpha$)
for J$^{\pi} = 0^{\pm}, 1^{\pm}, 2^{\pm}$ and $^{11}$Li (n+n+$^9$Li)
for $\frac{1}{2}^{\pm}$, $\frac{3}{2}^{\pm}$,
$\frac{5}{2}^{\pm}$. Ground state properties, strength functions,
Coulomb dissociation cross sections, phase shifts, complex $S$-matrix
poles are computed and compared to available experimental data. We
find enhancements of the strength functions at low energies and a
number of low-lying $S$-matrix poles.

{PACS numbers:  21.45.+v, 11.80.Jy, 21.10.Dr, 21.60.Gx}
\end{abstract}

\section{Introduction}

The present paper is part of a sequence discussing the general
properties of three-body halos
\cite{han95,tan96,rii94,han87,fed93b,zhu93}.  These papers all deal
with three-body systems, weakly bound and spatially extended compared
to the energy and range of the two-body interactions
\cite{fed94a,fed95,gar96,cob97}.  Most of the detailed information
about halo nuclei is obtained from reaction experiments
\cite{ber93,kor97,zin97,sac93,shi95}. Fragmentation reactions of
three-body halo nuclei were studied in the sudden approximation for
$^{11}$Li and $^{6}$He with special emphasis on the effects of final
state interactions, which in other words are the effects of the
two-body continuum \cite{gar96,bar93,kor94,zin95}.

Borromean systems, where no binary subsystem is bound, are
particularly interesting three-body halo candidates. They have by
definition a relatively low binding energy. The established nuclear
prototypes are $^{6}$He (n+n+$\alpha$) and $^{11}$Li (n+n+$^{9}$Li)
and other examples are expected further up along the neutron dripline.
The structure of $^{6}$He is fairly well understood whereas the
structure of $^{11}$Li is still controversial. The reason is
essentially the large amount of knowledge, respectively the lack of
knowledge, about the two-body subsystems.

The number of bound states for Borromean systems is almost always
limited to the ground state.  The effective two-body interactions must
be weak enough to exclude bound states and strong enough to bind the
three-body system. Therefore one or more two-body resonances must be
present at low energy. Then the low-lying continuum three-body
spectrum would inevitably have rather complicated structure. This has
strong implications for the analyses and the understanding of the data
accumulating from experiments with nuclear halos and Borromean nuclei
\cite{dan93a,dan93b,dan94,fer93,cso93,aoy95,dan97,cob97a}.

The three-body continuum problem has been the subject of numerous
investigations and tremendous progress has been achieved in recent
years \cite{glo96}.  However, a number of problems still remain
and unfortunately they include reliable computations of Borromean
continuum spectra \cite{fri95}.  Other difficult problems include the
continuum three-body Coulomb problem \cite{lin95} and scattering above
threshold of one particle on a two-body bound structure as for example
nucleon scattering on deuterons \cite{kie96,kol97}.  The problems can
crudely be divided into those dealing with the more technical aspects
like the accuracies of the employed analytical and numerical methods,
and physics issues addressing the behavior of the binary subsystems,
especially knowledge of the effective two-body interactions. It is
necessary, but not always easy, to distinguish between these
difficulties.  Different treatments are usually needed for short-range
and long-range interactions and for energies below or above possible
two-body thresholds. The concepts and problems described above are to
a large extent general and of interest in other subfields of physics
\cite{lin95,efi70,esr96,ric94,goy95,ric92,car93}. 

The general discussions of structure and break-up reactions of halo
nuclei should be extended to the three-body continuum. Specific
investigations are available, but imprecise
\cite{dan93a,dan93b,dan94,fer93}. The technical difficulties
formulated in coordinate space are related to the necessary
computation of the behavior of the effective potential at
large-distance. Fortunately, a method treating the large distances
analytically and the short distances numerically has recently become
available \cite{fed93,fed96}.  The method is very powerful in
structure calculations as demonstrated by the succesful investigation
of the Efimov effect \cite{fed95,fed93,fed94c,jen97}.  However, the
implementation has so far essentially concentrated on bound
structures, but generalization to applications in the three-body
continuum is straightforward.

The purpose of this paper is to (i) describe the details of a method
to compute low-energy three-body continuum spectra for particles with
or without intrinsic spin, (ii) derive asymptotic large-distance
expressions allowing simple computations of the corresponding
effective three-body potential for arbitrary angular momenta and
arbitrary short-range two-body potentials, (iii) apply the method in
detailed realistic numerical computations of the continuum structure
and various observables for $^{6}$He and $^{11}$Li. 

The paper generalizes first the analytic results obtained for $s$-waves
and square well potentials \cite{jen97}. Then the method is
applied to detailed studies of the continuum structure for the
Borromean nuclei $^{6}$He (n+n+$^4$He) and $^{11}$Li (n+n+$^9$Li).
Brief reports describing some  numerical results are available
in the literature \cite{cob97a,fed96}.

After the introduction we give in section 2 a general description of
the method and then we concentrate on two cases of special
interest. In section 3 and 4 we compute in details the properties of
$^{6}$He and $^{11}$Li, respectively. In section 5 we give a brief
summary and the conclusions. A convenient general expression for the
transformation between different sets of Jacoby coordinates is derived
in the appendix.

\section{Theory}
We shall consider a system of three interacting inert ``particles''.
Their intrinsic degrees of freedom are frozen and only the three-body
(halo) degrees of freedom shall be treated here. In this section we
first describe the general method of solving the Faddeev equations
using hyperspherical coordinates. In particular we specify the
boundary conditions at large distances by introducing the $S$-matrix
or equivalently the $R$-matrix. The angular equations at large
distances are then treated essentially analytically. We then consider
a system of two identical neutrons surrounding a core first with
finite spin and then with spin zero. Finally in this section we give
the expressions for strength functions and Coulomb dissociation cross
sections. We shall follow the method and the notation established
previously \cite{fed94a,fed95,gar96,fed93,fed96,cob97a}.

\subsection{Method}
 The k'th particle has mass $m_{k}$, charge $eZ_k$, coordinate ${\bf
r}_k$ and spin ${\bf s}_k$. The two-body interactions between the
particles $i$ and $j$ are $V_{ij}$.  We shall use the three sets of
Jacobi coordinates (${\bf x}_i,{\bf y}_i$) and the corresponding three
sets of hyperspherical coordinates $(\rho,\Omega_{i})$=($\rho$,
$\alpha_i$, $\Omega_{x_i}$, $\Omega_{y_i}$), see for example
\cite{zhu93,fed94a}.  The volume element in terms of one of the sets
of hyperspherical coordinates is given by $\rho^5{\rm d}\Omega{\rm
d}\rho$, where ${\rm d}\Omega=\sin^2 \alpha \cos^2 \alpha {\rm
d}\alpha {\rm d}\Omega_x {\rm d}\Omega_y$. The kinetic energy operator
is
\begin{eqnarray} \label{e10}
T=\frac{\hbar ^{2}}{2m}\left( -\rho ^{-5/2}\frac{\partial ^{2}}{\partial
\rho ^{2}}\rho ^{5/2}+\frac{15}{4\rho ^{2}}+\frac{\hat{\Lambda}^{2}}{\rho
^{2}}\right) \; , \\ \label{e15}
	\hat \Lambda^2=-{1 \over \sin(2\alpha)}
      {{\partial}^2\over {\partial} \alpha^2} \sin(2\alpha)  
	+{\hat l_x^2 \over {\sin^2 \alpha}}
 	+{\hat l_y^2 \over {\cos^2 \alpha}} -4  \; , 
\end{eqnarray}
where the angular momentum operators $\hat l_{x}^2$ and $\hat l_{y}^2$
are related to the ${\bf x}$ and ${\bf y}$ degrees of freedom and $m$
is a normalization mass arising from the definition of $\rho$. In the
following $m$ is assumed to be the nucleon mass.

The total wave function $\Psi_{J M}$ of the three-body system (with
total spin $J$ and projection $M$) is written as a
sum of three components $\psi^{(i)}_{ J M}$, which in turn for each $\rho$
are expanded on a complete set of generalized angular functions
$\Phi^{(i)}_{n}(\rho,\Omega_i)$
\begin{equation} \label{e20}
\Psi_{J M}= \sum_{i=1}^3 \psi^{(i)}_{ J M}(\mbox{\bf x}_i, \mbox{\bf y}_i)
= \frac {1}{\rho^{5/2}} 
  \sum_{n=1}^{\infty} f_n(\rho)  
\sum_{i=1}^3 \Phi^{(i)}_{n}(\rho ,\Omega_i) \; ,
\end{equation}
where $\rho^{-5/2}$ is the radial phase space factor.

The angular functions are now for each $\rho$ chosen as the
eigenfunctions of the angular part of the Faddeev equations:
\begin{equation} \label{e25}
{\hbar^2 \over 2m}\frac{1}{\rho^2} \left(\hat\Lambda^2-\lambda_n(\rho)\right)
\Phi^{(i)}_{n} 
 +V_{jk} (\Phi^{(i)}_{n}+\Phi^{(j)}_{n}+
                \Phi^{(k)}_{n}) = 0   \; ,
\end{equation}
where $\{i,j,k\}$ is a cyclic permutation of $\{1,2,3\}$.

The radial expansion coefficients $f_n(\rho)$ are obtained from a
coupled set of ``radial'' differential equations \cite{fed94a,kvi91},
i.e.
\begin{eqnarray} \label{e30}
   \left(-\frac{\partial ^{2}}{\partial \rho ^{2}} 
   -{2m(E-V_3(\rho))\over\hbar^2}+ \frac{1}{\rho^2}\left( \lambda_n(\rho) +
  \frac{15}{4}\right)  -Q_{n n} \right)f_n(\rho)
 \nonumber  \\
  = \sum_{n'\neq n}   \left( 2P_{n n'} \frac{\partial }{\partial \rho } 
   + Q_{n n'} \right)f_{n'}(\rho) \; ,
\end{eqnarray}
where $E$ is the three-body energy, $V_3(\rho)$ is an anticipated
additional three-body potential and the functions $P$ and $Q$ are
defined as angular integrals:
\begin{equation}\label{e40}
   P_{n n'}(\rho)\equiv \sum_{i,j=1}^{3}
   \int d\Omega \Phi_n^{(i)\ast}(\rho,\Omega)
   {\partial\over\partial\rho}\Phi_{n'}^{(j)}(\rho,\Omega)  \; ,
\end{equation}
\begin{equation} \label{e45}
   Q_{n n'}(\rho)\equiv \sum_{i,j=1}^{3}
   \int d\Omega \Phi_n^{(i)\ast}(\rho,\Omega)
   {\partial^2\over\partial\rho^2}\Phi_{n'}^{(j)}(\rho,\Omega)  \; .
\end{equation}
The diagonal part of the effective potential is then
\begin{equation} \label{e44}
  \frac{\hbar^2}{2m} \left( ( \lambda_n(\rho) +
  \frac{15}{4}) \rho^{-2}  - Q_{n n} \right) + V_3(\rho) \; .
\end{equation}

For Borromean systems the coupling terms $P$ and $Q$ approach zero at
least as fast as $\rho^{-3}$ \cite{fed94a,fed93,jen97}.  We can then
choose those solutions $\Psi_{n'}$ to eq.(\ref{e20}) where the
large-distance ($\rho \rightarrow \infty)$ boundary conditions for
$f^{(n')}_{n}$ are given by \cite{tay72}
\begin{equation} \label{e50}
f^{(n')}_n(\rho) \rightarrow \delta_{n,n'} 
 F^{(-)}_{n}(\kappa \rho) - S_{n,n'} F^{(+)}_{n}(\kappa \rho) \; .
\end{equation}
The $S$-matrix introduced here is a unitary matrix,
$\kappa^2=2mE/\hbar^2$ and $F^{(\pm)}_{n}$ are related to the Hankel
functions of integer order by
\begin{eqnarray} \label{e55}
F^{(\pm)}_{n}(\kappa \rho) = 
 \sqrt{\frac{m \rho}{4 \hbar^2}}\, H^{(\pm)}_{K_{n+2}}(\kappa \rho)  
 \rightarrow \sqrt{\frac{m}{2 \pi \kappa \hbar^2}} 
\exp\left[\pm i\kappa \rho \pm 
 {i\pi\over 2}(K_n+{3\over 2})\right] \; ,
\end{eqnarray} 
where $K_n$ is the hyperspherical quantum number corresponding to the
value $K_n\times (K_n+4)$ approached at large distance by the angular
eigenvalue $\lambda_n$.  The continuum wave functions $\Psi_{JM}$ are
orthogonal and normalized to delta functions in energy. Sometimes it
is more convenient to work with the $R$-matrix given as $R= i
(1-S)/(1+S)$. The boundary conditions must then be changed into
$\sin$ and $\cos$ instead of the exponentials in eq.(\ref{e55}).

By diagonalization of the $S$- (or $R$)-matrix we obtain
eigenfunctions and eigenphases. These phase shifts reveal the
continuum structure of the system. In particular, a rapid variation
with energy indicates a resonance.  A precise computation of
resonances and related widths can be done by use of the complex energy
method, where eq.(\ref{e30}) is solved for $E=E_r-i\Gamma/2$ with the
boundary condition $f_n \propto \sqrt{\frac{m \rho}{4 \hbar^2}}\,
H^{(+)}_{K_{n+2}}(\kappa \rho)$.  These solutions correspond to poles
of the $S$-matrix \cite{tay72}.

\subsection{Angular eigenvalue equation}
The angular functions $\Phi^{(i)}_{n}(\rho ,\Omega_i)$ are expanded in
products of the three-body spin functions $\chi_{s_x s_y S m_s}^{(i)}$
and spherical harmonics $Y_{\ell_x m_x}(\Omega_{x_i})$ and $Y_{\ell_y
m_y}(\Omega_{y_i})$.  The orbital angular momenta and their
projections associated with {\bf x} and {\bf y} are $(\ell_x, m_x)$
and $(\ell_y, m_y)$ while the spins of the two particles connected by
the {\bf x} coordinate couple to the spin $s_x$, which coupled to the
spin $s_y$ of the third particle results in the total spin $S$ and its
projection $m_s$. Indicating these angular momentum couplings by
$\otimes$ the result can be written
\begin{eqnarray} \label{e60}
\Phi^{(i)}_{n}(\rho ,\Omega_i) = \sum_{\ell_x \ell_y L s_x S}
\frac{\phi_{n \ell_x \ell_y L s_x S}^{(i)}(\rho, \alpha_i)}{\sin(2\alpha_i)}
\left[ Y_{\ell_x \ell_y}^{LM_L}(\Omega_{x_i},\Omega_{y_i})
 \otimes \chi_{s_x s_y S m_s}^{(i)} \right]^{J M}  \; ,
\end{eqnarray} 
where $\sin(2\alpha_i)$ is a factor related to phase space and 
\begin{eqnarray} \label{e61}
Y_{\ell_x \ell_y}^{LM_L}(\Omega_{x_i},\Omega_{y_i}) \equiv
\left[Y_{\ell_x m_x}(\Omega_{x_i}) \otimes Y _{\ell_y m_y}(\Omega_{y_i})
\right]^{LM_L} \; ,
\end{eqnarray} 
where the projections of the intermediate couplings are given although
the final result is independent of them.

To solve the angular Faddeev equations the components in
eq.(\ref{e25}) must be expressed in one Jacobi coordinate set, say
labeled by $i$.  The wave functions $\phi_{n \ell_x \ell_y L s_x
S}^{(j) }(\rho, \alpha_j)/\sin(2\alpha_j)$, which only depend on
$\alpha _j$ and $\rho$, are first expressed in terms of the i'th set
of hyperspherical coordinates. The equations are multiplied from the
left by the square bracket in eq.(\ref{e60}) and subsequently
integrated over the four angular variables describing the directions
of ${\bf x}_i$ and ${\bf y}_i$. The interaction $V_{jk}$, only
depending on the distance between the particles, is independent of
these angles.

The operator describing this transformation from the $j$'th to the
$i$'th Jacobi coordinate system is denoted $R_{ij}$. This operation
maintains both total spin and total orbital angular momentum. The
result of the transformation from a specific set of angular momentum
states $\ell_x^\prime \ell_y^\prime L$ is projected on the set $\ell_x
\ell_y L$. This operator $R_{ij}^{\ell_x \ell_y \ell_x^\prime
\ell_y^\prime L}$ is then given by
\begin{eqnarray}\label{e75}
R_{ij}^{\ell_x \ell_y \ell_x^\prime \ell_y^\prime L}
\left[\frac{\phi_{n \ell_x^\prime \ell_y^\prime L s_x^\prime S}^{(j) }(\rho,
\alpha_j)}{\sin(2\alpha_j)}\right]  \equiv
\int {\rm d}\Omega_{x_i} {\rm d}\Omega_{y_i} 
\left[Y_{\ell_x \ell_y}^{LM_L}(\Omega_{x_i},\Omega_{y_i})\right]^*
 \nonumber \\
\times \frac{\phi_{n \ell_x^\prime \ell_y^\prime L s_x^\prime S}^{(j) }(\rho,
\alpha_j)}{\sin(2\alpha_j)}
Y_{\ell_{x}^\prime \ell_{y}^\prime}^{LM_L}(\Omega_{x_j},\Omega_{y_j}) \; .
\end{eqnarray}

When the two-body interaction is assumed to be diagonal in the total
two-body spin we now rewrite the angular eigenvalue equation in
eq.(\ref{e25}) as
\begin{eqnarray} \label{e65}
& \left(-\frac{\partial^2 } {\partial \alpha_i^2} + 
\frac{\ell_x(\ell_x+1)}{\sin^2 \alpha_i} 
+ \frac{\ell_y(\ell_y+1)}{\cos^2 \alpha_i} 
+ \rho^2 v^{s_xS}_i(\rho \sin{\alpha_i}) - \nu^2_n(\rho)\right)
 \phi_{n \ell_x \ell_y L s_x S}^{(i) }(\rho, \alpha_i)
  \nonumber \\ 
&  + \rho^2 {\sin(2\alpha _i)}  v^{s_xS}_i(\rho \sin{\alpha_i})
 \sum_{\ell_x^\prime \ell_y^\prime s_x^\prime}
\left(C_{s_x s_x{^\prime} S}^{ij} 
R_{ij}^{\ell_x \ell_y \ell_x^\prime \ell_y^\prime L} 
 \left[\frac{\phi_{n \ell_x^\prime \ell_y^\prime L s_x^\prime S}^{(j) }(\rho,
\alpha_j)}{\sin(2\alpha_j)}\right]  
   \right.  \nonumber \\ 
& \left.
 +  C_{s_x s_x{^\prime} S}^{ik} 
R_{ik}^{\ell_x \ell_y \ell_x^\prime \ell_y^\prime L}
\left[\frac{\phi_{n \ell_x^\prime \ell_y^\prime L s_x^\prime S}^{(k) }(\rho,
\alpha_k)}{\sin(2\alpha_k)} \right] \right) = 0  \; ,
\end{eqnarray}
where $\nu^2_n(\rho) \equiv \lambda_n(\rho)+4$ and the reduced and spin
averaged interactions are given by
\begin{equation}\label{e68}
v^{s_xS}_i(x)= \langle  \chi_{s_x s_y S m_s}^{(i)} |  \frac{2m}{\hbar^2}
 V_{jk}(\frac{x}{\mu_{jk}}) | \chi_{s_x s_y S m_s}^{(i)} \rangle  \; 
\end{equation}
with $m\mu^2_{jk}\equiv m_jm_k/(m_j+m_k)$. The coefficients $C_{s_x
s_x{^\prime} S}^{ik}$, expressing the overlap of the spin functions, are
defined by
\begin{equation}\label{e70}
 C_{s_x s_x{^\prime} S}^{ik} = \langle \chi_{s_x s_y S m_s}^{(i)} |
 \chi_{s_x^\prime s_y S m_s}^{(k)} \rangle \; .
\end{equation}
These matrix elements are independent of $m_s$, symmetric, i.e.\ $C_{s
s{^\prime} S}^{ik} = C_{s{^\prime} s S}^{ki}$ and diagonal in $s_x$
and $s_x^\prime$ for $i=k$, i.e. $C_{s s^\prime S}^{ii} =
\delta_{ss^\prime}$.

\subsection{Large-distance angular eigenvalues}
For large $\rho$ only small $\alpha$-values contribute in
eq.(\ref{e65}) to the terms proportional to $\rho^2 v^{s_xS}_i(\rho
\sin{\alpha_i})$. These potentials are assumed to have short ranges
and they vanish consequently for large $\rho$ for all $\alpha_i$
except in a narrow region around zero. We assume that they vanish
exponentially or at least as fast as $1/\rho^3$, for distances
larger than the ranges of the potentials
\cite{fed94a,fed93,jen97}. The two terms described by the
transformations $R_{ij}$ and $R_{ik}$ in eq.(\ref{e65}) can then be
approximated by their expansion to leading order in the variable
$\alpha_i$.

We show in appendix A that all partial waves decouple to leading order
in $\alpha_i$ or in $1/\rho$ with the essential exception of $s$-waves
in the $x$-degree of freedom, i.e. the components with $\ell_{x_i}=0,
i=1,2,3$ and the total orbital angular momentum $L=\ell_{y_i}$.  This
means that an expansion in powers of $\alpha_i$ of the terms obtained
from the transformation $R_{ij}$ only provides non-zero contributions
in the limit of $\alpha_i=0$ for $\ell_x=\ell_x^\prime=0$,
$\ell_y=\ell_y^\prime=L$. These finite contributions are for
$\alpha_i=0$ found to be
\begin{eqnarray}\label{e80}
R_{ij}^{0 L 0 L L}
\left[\frac{\phi_{n 0 L L s_x^\prime S}^{(j) }(\rho,
\alpha_j)}{\sin(2\alpha_j)}\right] =  (-1)^L   
\frac{\phi_{n 0 L L s_x^\prime S}^{(j) }(\rho, \varphi_{k})} 
{\sin(2 \varphi_{k})} \; , \\  \label{e83}
 \tan \varphi_{k} = \sqrt{\frac{m_k(m_i+m_j+m_k)}{m_i m_j}} \; ,
\end{eqnarray}
where $\{i,j,k\}$ again must be a permutation of $\{1,2,3\}$. Higher
order terms in $\alpha_i$ are neglected.  Non-zero $\ell_{x}$-values
had produced leading terms of at least first order in $\alpha_i$ in the
expression analogous to eq.(\ref{e80}).

Thus for non-zero $\ell_{x}$-values the angular eigenvalue equations
in eq.(\ref{e65}) decouple asymptotically and reduce to
\begin{eqnarray} \label{e85}
  \left(-\frac{\partial^2 } {\partial \alpha_i^2} + 
\frac{\ell_x(\ell_x+1)}{\sin^2 \alpha_i} 
+ \frac{\ell_y(\ell_y+1)}{\cos^2 \alpha_i} \right. \nonumber \\  
 \left. 
+ \rho^2 v^{s_xS}_i(\rho \sin{\alpha_i})  - \nu^2_n(\rho)\right)
  \phi_{n \ell_x \ell_y L s_x S}^{(i) }(\rho, \alpha_i) = 0 \; 
\end{eqnarray}
for all sets of values of $\ell_{x} \neq 0, \ell_{y}, L, s_x, S$.  The
large-distance asymptotic eigenvalues $\nu^2_n=(K+2)^2$ related to
these partial waves approach the hyperspherical spectrum, where $K$ is
odd or even natural numbers depending on the parity.  This asymptotic
behavior is reached on a distance scale defined by the short range of
the interactions $v^{s_xS}_i$ in eq.(\ref{e85}).

For $\ell_{x}=0$ insertion of eq.(\ref{e80}) into eq.(\ref{e65})
gives instead the three coupled asymptotic angular equations
\begin{equation} \label{e90}
 \left(\frac{\partial^2 } {\partial \alpha_i^2} 
+ \kappa_i^2(\rho,\alpha_i) \right)
  \phi_{n 0 L L s_x S}^{(i) }(\rho, \alpha_i)
     =  2\alpha _i (-1)^L   C^{(i)}_{Ls_xS}
  \rho^2 v^{s_xS}_i(\rho \sin{\alpha_i}) \; ,
\end{equation}
\begin{equation} \label{e92}
\kappa_i^2(\rho,\alpha_i) = - \left[\frac{L(L+1)}{\cos^2 \alpha_i}
+ \rho^2 v^{s_xS}_i(\rho \sin{\alpha_i}) - \nu^2_n(\rho)\right] \; ,
\end{equation}
\begin{equation} \label{e95}
 C^{(i)}_{Ls_xS} \equiv  \sum_{s_x^\prime} \left(C_{s_x s_x{^\prime} S}^{ij} 
 \frac{\phi_{n 0 L L s_x^\prime S}^{(j) }(\rho, \varphi_{k})} 
{\sin(2 \varphi_{k})}
 +  C_{s_x s_x{^\prime} S}^{ik} 
\frac{\phi_{n 0 L L s_x^\prime S}^{(k) }(\rho, \varphi_{j})} 
{\sin(2 \varphi_{j})}
 \right) \; .
\end{equation}
Also these eigenvalue solutions $\nu^2_n$ converge towards the
hyperspherical spectrum as $\rho$ increases. However, due to the
coupling the asymptotic values are now approached over a distance
defined by the scattering lengths, which might be very much larger
than the ranges of the interactions.

As mentioned above the potentials $\rho^2 v^{s_xS}_i(\rho
\sin{\alpha_i})$ vanish for large $\rho$ for all $\alpha_i$ except in
a narrow region around zero. The conditions for the effective range
approximation therefore become better and better fulfilled as $\rho$
increases and any potential with the same scattering length and
effective range would lead to the same results.  Let us then in the
region of large $\rho$ use square well potentials $V_{jk}(r) = -
S^{(i)}_0(s_js_k) \Theta(r<R^{s_xS}_i)$, or equivalently expressed by
the reduced quantities $v^{s_xS}_{i}(x) = - s^{(i)}_0(s_xS)
\Theta(x<X^{s_xS}_i=R^{s_xS}_i \mu_{jk})$, where the range and depth
parameters are adjusted to reproduce the two-body scattering lengths
and effective ranges of the initial potential. The corresponding
solutions are then accurate approximations to our original problem at
distances larger than $2R^{s_xS}_i$ \cite{jen97}.

The square well potentials $v^{s_xS}_i(\rho \sin{\alpha_i})$ are zero
in region II defined by $\alpha_i>\alpha^{(i)}_0({s_xS}) =
\arcsin(X^{s_xS}_i/\rho)$. Then eq.(\ref{e90}) is especially simple,
i.e.
\begin{equation} \label{e72}
  \left(
 - \frac{\partial^2}{\partial \alpha_i^2} + \frac{L(L+1)}{\cos^2\alpha_i}
 -\nu^2_n  
  \right) \phi_{n0LLs_xS}^{(i)}(\rho,\alpha_i) = 0 \; 
\end{equation}
and the solutions, vanishing at $\alpha_i=\pi/2$, are given by
\begin{eqnarray}\label{e76}
 \phi_{n0LLs_xS}^{(i,II)}(\rho,\alpha) = A_{n0LLs_xS}^{(i)} 
 P_{L}(\nu_n,\alpha) \; , \label{e82}  \\
 P_{L}(\nu_n,\alpha) \equiv \cos^L\alpha
 \left(
 \frac{\partial}{\partial \alpha} \frac{1}{\cos\alpha}
\right)^L \sin\left[ \nu_n 
\left(\alpha - \frac{\pi}{2} \right) \right] \; 
\end{eqnarray}
for arbitrary constants $A_{n0LLs_xS}^{(i)}$.

The potentials $v^{s_xS}_i(\rho \sin{\alpha_i})$ are finite and
constant for large $\rho$ in region I defined by $\alpha_i <
\alpha^{(i)}_0({s_xS}) \ll 1$. Then eq.(\ref{e90}) is approximately
\begin{eqnarray}\label{e86}
  \left(
  \frac{\partial^2}{\partial \alpha_i^2}  + \kappa_i^2(\rho,\alpha_i=0) 
  \right) \phi_{n0LLs_xS}^{(i)}(\rho,\alpha_i) = 
 - 2 \alpha_i (-1)^L  \rho^2 s_0^{(i)}({s_xS})  C^{(i)}_{Ls_xS} \; ,
\end{eqnarray}
where the wave functions in $C^{(i)}_{Ls_xS}$ in eq.(\ref{e95}) must
be $\phi^{(i,II)}_{n0LLs_xS}$.  The solutions to eq.(\ref{e86}),
vanishing for $\alpha_i=0$, are then
\begin{eqnarray} \label{e89}
\phi_{n0LLs_xS}^{(i,I)}(\rho,\alpha) = 
B_{n0LLs_xS}^{(i)}\sin(\alpha \kappa_i(\rho,\alpha=0)) 
- 2 \alpha (-1)^L   
\frac{\rho^2 s_0^{(i)}({s_xS})}{\kappa_i^2(\rho,\alpha=0)}  C^{(i)}_{Ls_xS} \; 
\end{eqnarray}
for arbitrary constants $B_{n0LLs_xS}^{(i)}$, where $\kappa_i$ are
defined in eq.(\ref{e92}).

Matching the solutions, eqs.(\ref{e76}) and (\ref{e89}), and their
derivatives at $\alpha_i=\alpha^{(i)}_0({s_xS})$ gives a linear set of
equations for $A_{n0LLs_xS}^{(i)}$ and $B_{n0LLs_xS}^{(i)}$ with given
$L$ and $S$ for $i=1,2,3$ and all possible $s_x$. Physical solutions
are then only obtained when the corresponding determinant is zero.
This is the quantization condition for $\nu^2$ (or $\lambda$) and as
such the eigenvalue equation determining the large-distance asymptotic
behavior of $\lambda(\rho)$.

The square well solution in eq.(\ref{e89}) is not exact since the
first order expansion in $ \alpha_i$ is used in eq.(\ref{e90}) and
(\ref{e86}) and consequently also in the last term of
eq.(\ref{e89}). Improvements could be obtained by using eq.(\ref{e76})
in eq.(A\ref{a31}) and thereby changing the right hand sides of
eqs.(\ref{e80}), (\ref{e90}), (\ref{e86}) and (\ref{e89}). For $L=0$
these expressions are given in \cite{jen97}.

Also the eigenvalue equation for non-zero $\ell_x$-values in
eq.(\ref{e85}) can be solved for square well potentials. For
$\alpha_i>\alpha^{(i)}_0({s_xS})$ in region II, we have the equation
\begin{equation} \label{e100}
  \left(
 - \frac{\partial^2}{\partial \alpha_i^2} + 
\frac{\ell_{x}(\ell_{x}+1)}{\sin^2\alpha_i} + 
\frac{\ell_{y}(\ell_{y}+1)}{\cos^2\alpha_i} -\nu^2_n  
  \right) \phi_{n\ell_{x}\ell_{y}Ls_xS}^{(i)}(\rho,\alpha_i) = 0 \; 
\end{equation}
and the solutions vanishing at $\pi/2$ are then
\begin{equation} \label{e105}
\phi_{n\ell_{x}\ell_{y}Ls_xS}^{(i,II)}(\rho,\alpha) = 
A_{n\ell_{x}\ell_{y}Ls_xS}^{(i)} N_{n\ell_x\ell_y}
 \sin^{\ell_{x}} \alpha \cos^{\ell_{y}} \alpha 
P^{\ell_{x}+1/2,\ell_{y}+1/2}_n(\cos(2\alpha))
\end{equation}
for arbitrary constants $A_{n\ell_{x}\ell_{y}Ls_xS}^{(i)}$, where
$P^{\ell_{x}+1/2,\ell_{y}+1/2}_n$ are the Jacobi polynomials and
$N_{n\ell_x\ell_y}$ are normalization constants given explicitly in
appendix A.

For $\alpha_i<\alpha^{(i)}_0({s_xS})\ll 1$ in region I eq.(\ref{e85})
is approximately
\begin{equation} \label{e110}
  \left(
 - \frac{\partial^2}{\partial \alpha_i^2} + 
\frac{\ell_{x}(\ell_{x}+1)}{\alpha_i^2} - \kappa_i^2(\rho,\alpha=0)  \right) 
\phi_{n\ell_{x}\ell_{y}Ls_xS}^{(i)}(\rho,\alpha_i) = 0 \; 
\end{equation}
with the solutions vanishing as $\alpha_i^{\ell_{x}+1}$ at $\alpha_i=0$
\begin{equation} \label{e120}
\phi_{n\ell_{x}\ell_{y}Ls_xS}^{(i,I)}(\rho,\alpha) = 
B_{n\ell_{x}\ell_{y}Ls_xS}^{(i)} j_{\ell_{x}}(\alpha \kappa_i(\rho,\alpha=0)) 
\end{equation}
\begin{equation} \label{e112}
\kappa_i^2(\rho,\alpha=0) = - \left[\ell_{y}(\ell_{y}+1)
- \rho^2 s_0^{(i)}({s_xS}) - \nu^2_n \right] \; ,
\end{equation}
where $B_{n\ell_{x}\ell_{y}Ls_xS}^{(i)}$ is an arbitrary constant and
$j_{\ell_x}$ is the spherical Bessel function, i.e. the usual solution
to the radial two-body Schr\"{o}dinger equation for an angular
momentum $\ell = \ell_{x}$.

Matching the logarithmic derivatives of the solutions in
eqs.(\ref{e105}) and (\ref{e120}) then provides the quantization
condition for $\nu^2$ (or $\lambda$) and therefore the large-distance
asymptotic behavior of $\lambda(\rho)$.

\subsection{Large-distance behavior for two neutrons and a core}
We shall now consider a system of two neutrons (labeled 2,3) and a
core (labeled 1) with spin $s_c$.  For a given total spin and for each
set of orbital quantum numbers the six possible components,
$\phi_{n\ell_{x}\ell_{y}Ls_xS}^{(i)}, i=1,2,3,$ each with two
$s_x$-values, are related to the three-body spin wave functions
$\chi_{s_x=0,1}^{(1)}$, $\chi_{s_x=s_c\pm1/2}^{(2)}$,
$\chi_{s_x=s_c\pm1/2}^{(3)}$, where the first set of Jacobi
coordinates corresponds to the $x$-coordinate between the two
neutrons. Due to the Pauli principle only three of these wave
functions $\phi$ are independent and the remaining components are
determined by antisymmetry, i.e.
\begin{eqnarray} \label{e125} 
\phi_{n\ell_{x}\ell_{y}Ls_xS}^{(1)} &=& 0 \; \; {\rm for \; \; odd} \; \; 
\ell_{x} +s_x \; , \\
\phi_{n\ell_{x}\ell_{y}Ls_xS}^{(3)} &=& (-)^{s_c+1/2-s_x+\ell_x-1}
\phi_{n\ell_{x}\ell_{y}Ls_xS}^{(2)} \; .
\end{eqnarray}

Specifically the three independent $s$-wave components can be
characterized by $L$, $S$ and one value of $s_x$, i.e.
$\phi_{LS}^{(1)} \equiv \phi_{n0LL0S}^{(2)}$, $\phi_{LS}^{(2)} \equiv
\phi_{n0LLs_c-1/2S}^{(2)}$, $\phi_{LS}^{(3)} \equiv
\phi_{n0LLs_c+1/2S}^{(3)}$. These components are coupled over a
distance defined by the scattering lengths whereas all other partial
waves decouple for large $\rho$ above a distance scale defined by the
range of the interactions.

The components $\phi_{LS}^{(i)}$ obey for large $\rho$ the coupled
angular Faddeev equations in eq.(\ref{e90}), where the coefficients
$C^{(i)}_{LS} \equiv C^{(i)}_{Ls_{x_i}S}$ ($s_{x_1}=0,
s_{x_2}=s_c-\frac{1}{2}, s_{x_3}=s_c+\frac{1}{2}$) explicitly are
given by
\begin{eqnarray} \label{e130}
& C^{(1)}_{LS}=2 C^{12}_{0,s_c-1/2,S}
 \frac{\phi^{(2)}_{LS}(\varphi)}{\sin(2 \varphi)} -2
 C^{12}_{0,s_c+1/2,S} \frac{\phi^{(3)}_{LS}(\varphi)} {\sin(2
 \varphi)} & \\ 
& C^{(2)}_{LS}=C^{12}_{0,s_c-1/2,S}
 \frac{\phi^{(1)}_{LS}(\varphi)}{\sin(2 \varphi)} +
 C^{23}_{s_c-1/2,s_c-1/2,S} \frac{\phi^{(2)}_{LS}(\tilde
 \varphi)}{\sin(2 \tilde \varphi)} + C^{23}_{s_c-1/2,s_c+1/2,S}
 \frac{\phi^{(3)}_{LS}(\tilde \varphi)} {\sin(2 \tilde \varphi)} &
 \; \nonumber \\ 
& C^{(3)}_{LS}=C^{13}_{0,s_c+1/2,S}
 \frac{\phi^{(1)}_{LS}(\varphi)}{\sin(2 \varphi)} +
 C^{23}_{s_c-1/2,s_c+1/2,S} \frac{\phi^{(2)}_{LS}(\tilde
 \varphi)}{\sin(2 \tilde \varphi)} - C^{23}_{s_c+1/2,s_c+1/2,S}
 \frac{\phi^{(3)}_{LS}(\tilde \varphi)} {\sin(2 \tilde \varphi)} &
 \nonumber ,
\end{eqnarray}
where we omitted the argument $\rho$ in the functions $\phi$ and further
defined  $\varphi = \varphi_{2} = \varphi_{3}$, $\tilde \varphi =
\varphi_{1}$.  For $S=s_c$ all terms are present, but for $S=s_c \pm
1$ the first term in $C^{(i)}_{LS}$ should be removed together with
the equation corresponding to $i=1$.

The spin overlap coefficients are explicitly given by
\begin{eqnarray}\label{e135} 
 C^{12}_{0,s_c-1/2,s_c} = C^{13}_{0,s_c-1/2,s_c} = 
- \sqrt{\frac{s_c}{2s_c+1}}  \\
 C^{12}_{0,s_c+1/2,s_c} = - C^{13}_{0,s_c+1/2,s_c}=
 \sqrt{\frac{s_c+1}{2s_c+1}}  \\
   C^{23}_{s_c-1/2,s_c+1/2,s_c} = \frac{\sqrt{4s_c(s_c+1)}}{2s_c+1} \\
 C^{23}_{s_c-1/2,s_c-1/2,s_c} =  C^{23}_{s_c+1/2,s_c+1/2,s_c} = 
- \frac{1}{2s_c+1} \\
C^{23}_{s_c-1/2,s_c-1/2,s_c + 1} =  C^{23}_{s_c+1/2,s_c+1/2,s_c - 1} =
 C^{23}_{s_c-1/2,s_c+1/2,s_c \pm 1} = 0  \\
 C^{23}_{s_c-1/2,s_c-1/2,s_c - 1} =  C^{23}_{s_c+1/2,s_c+1/2,s_c + 1} = 1
\end{eqnarray}

The potentials $\rho^2 v_i(\rho \sin{\alpha})$ approach for
sufficiently large $\rho$ the zero-range potentials, where the
sensitivity to the shape disappears. Any potential with the same
scattering length and effective range would then lead to results
accurate to the order $\rho^{-2}$. We shall therefore for convenience
use such equivalent square well potentials, where the solutions to
eq.(\ref{e90}) then again are given by eqs.(\ref{e76}) and
(\ref{e89}), and the large-distance physical solutions are obtained as
described above.

We shall now consider a system of two neutrons and a core with spin
$s_c=0$. All quantities where $s_c-\frac{1}{2}$ appears as an index
should now be substituted by zero.  Then the three coupled $s$-wave
equations in eq.(\ref{e90}) reduce for $S=0$ to two as seen from
eq.(\ref{e130}) where $C^{(2)}_{L0}$ then is zero.  These equations
are to leading order in $\alpha$ (large $\rho$) explicitly given by
\begin{eqnarray} \label{e140}
  \left(
 - \frac{\partial^2}{\partial \alpha_1^2} + \frac{L(L+1)}{\cos^2\alpha_1}
 + \rho^2 v_{\mbox{\scriptsize NN}}(\rho \sin{\alpha_1}) - \nu^2
   \right) \\  \nonumber 
 \times \phi_{L}^{(1)}(\rho,\alpha_1) = 
- 2 \alpha_1 (-1)^L   \rho^2 v_{\mbox{\scriptsize NN}}
(\rho \sin{\alpha_1})  C^{(1)}_{L} \; , \\ \label{e141}
  \left(
 - \frac{\partial^2}{\partial \alpha_3^2} + \frac{L(L+1)}{\cos^2\alpha_3}
 + \rho^2 v_{\mbox{\scriptsize N}c}(\rho \sin{\alpha_3}) - \nu^2
   \right) \label{e142}  \\  \nonumber 
 \times \phi_{L}^{(3)}(\rho,\alpha_3) =
- 2 \alpha_3 (-1)^L 
\rho^2 v_{\mbox{\scriptsize N}c}(\rho \sin{\alpha_3})  C^{(3)}_{L} \; ,
\end{eqnarray}
where $v_{\mbox{\scriptsize NN}}(x_1) = v_1^{(00)}(x_1)$,
$v_{\mbox{\scriptsize N}c}(x_2) = v_2^{(1/2,0)}(x_2) =
v_3^{(1/2,0)}(x_2)$, $\phi_{L}^{(i)} \equiv \phi_{L0}^{(i)}$,
$C_{L}^{(i)} \equiv C_{L0}^{(i)}$ and
\begin{equation}\label{e69}
   C^{(1)}_{L} = 2  \frac{\phi^{(3)}_{L}(\varphi)}{\sin(2 \varphi)}
 \; \;  \; , \; \;  \; 
 C^{(3)}_{L} =  \frac{\phi^{(1)}_{L}(\varphi)}{\sin(2 \varphi)}
 + \frac{\phi^{(3)}_{L}(\tilde{\varphi})}
{\sin(2 \tilde{\varphi})} \; .
\end{equation}
The equivalent square well solutions are again given by
eqs.(\ref{e76}) and (\ref{e89}), and the large-distance asymptotic
behavior are obtained as described above.

For $S=1$ only eq.(\ref{e142}) remains for $s$-waves now with $
C^{(3)}_{L} = \phi^{(3)}_{L}(\tilde{\varphi})/\sin(2
\tilde{\varphi})$. The square well solution and the large-distance
behavior are then easily obtained.

\subsection{Strength functions and Coulomb cross sections}
The strength functions $\frac{dB_{E\lambda}}{dE}$ describing electric
multipole excitations of the ground state $|J_0^{\pi_0}\rangle$ into
the continuum state $|nJ^{\pi}E\rangle$ are defined by
\begin{eqnarray}\label{e43}
 \frac{dB_{E\lambda}(E)}{dE} = \frac{1}{2J_0 + 1}
 \sum_{nJ^{\pi}} \left| \langle n J^{\pi} E ||
 M(E\lambda) || J_0^{\pi_0}  \rangle \right|^2 \;  , \\ \label{e47}
 M(E\lambda,\mu) = \sum_{k=1}^3 eZ_k r_k^\lambda Y_{\lambda \mu}(\hat r_k)\; .
\end{eqnarray} 
in terms of the reduced matrix element and the electrical multipole
operator $M$.  The corresponding sum rule is
\begin{eqnarray}\label{e48}
 \int {\rm d}E \frac{dB_{E\lambda}(E)}{dE} = \frac{2\lambda +1}{4\pi} 
 \sum _{k=1}^3 e^2Z_k^2
 \langle J_0^{\pi_0} | r_k^{2\lambda} | J_0^{\pi_0} \rangle \; ,
\end{eqnarray} 
where only the core contributes for a system of two neutrons around a
core.

Nuclear excitations of monopole type are possible with the
corresponding operator $\rho^2 = \sum_k ({\bf r}_k-{\bf R_c})^2$,
where ${\bf R}_c$ is the coordinate of the center of mass. The related
strength function $\frac{dN_{E0}}{dE}$ and the sum rule are then
\begin{eqnarray}\label{e145}
 \frac{dN_{E0}(E)}{dE}  = \frac{1}{2J_0 + 1}
 \sum_{n} \left| \langle n J_0^{\pi_0} E | \rho^2
 | J_0^{\pi_0}  \rangle \right|^2 \;  , \\ \label{e147}
 \int {\rm d}E \frac{dN_{E0}(E)}{dE}  = \frac{1}{4\pi} 
 \left(\langle J_0^{\pi_0} | \rho^4 | J_0^{\pi_0} \rangle 
 - \langle J_0^{\pi_0} | \rho^2 | J_0^{\pi_0} \rangle ^2 \right)\; ,
\end{eqnarray}

The Coulomb dissociation cross section can now be computed in the high
beam energy limit where the approximation of straight-line
trajectories is valid and only one photon is exchanged between
projectile and target.  The cross section is then obtained by
multiplying the electromagnetic transition matrix elements
$\frac{dB_{E\lambda}(E)}{dE}$ from ground state
to continuum states with the virtual photon spectrum
$n_{E\lambda}(\omega)$, which is given by \cite{ber88}
\begin{eqnarray}\label{e62}
 n_{E1}(\omega) = \frac{2}{\pi} Z_t^2 \alpha \frac{c^2}{v^2} \left [
 \xi K_0(\xi) K_1(\xi) - \frac{v^2\xi^2}{2c^2}(K_1(\xi)^2 - K_0(\xi)^2)\right]
 \; ,\\
n_{E2}(\omega)= \frac{2}{\pi} Z_t^2 \alpha \frac{c^4}{v^4} \left[
2 \left(1-\frac{v^2}{c^2}\right)K_1(\xi)^2 + 
\xi \left(2-\frac{v^2}{c^2}\right)^2 K_0(\xi) K_1(\xi)
\nonumber \;  \right.  \\ \left.
 - \frac{v^4\xi^2}{2c^4}(K_1(\xi)^2 - K_0(\xi)^2)\right] 
\end{eqnarray} 
and the resulting differential cross section is
\begin{eqnarray}\label{e57}
 \frac{{\rm d} \sigma_{E1}(E)}{ {\rm d} E} = \frac{16\pi^3\alpha}{9} 
n_{E1}(E^*/\hbar) \frac{1}{e^2} \frac{dB_{E1}(E^*)}{dE^*} \; , \\
\frac{{\rm d} \sigma_{E2}(E)}{ {\rm d} E} = \frac{4\pi^3\alpha}{75} 
\left( \frac{E^*}{\hbar c} \right)^2
n_{E2}(E^*/\hbar) \frac{1}{e^2} \frac{dB_{E2}(E^*)}{dE^*} \; ,
\end{eqnarray} 
where $K$ is the modified Bessel function, $\alpha = e^2/ \hbar c$,
$\xi=\frac{\omega R}{\gamma v}$, $v$ is the beam velocity,
$\gamma=1/\sqrt{1-v^2/c^2}$, $Z_t$ is the charge of the target and
$E^*=\hbar \omega = E_f-E_i$, where the final and the initial energies
are labeled by f and i. The dipole is usually by far the largest
contributor. In any case for most halo nucei (the quadrupole
excitation for $^6$He is an exception) the information about
higher-lying angular momentum states is not experimentally available
and very difficult to predict theoretically due to the lack of
knowledge about the binary subsystems.

\section{The $^{6}$He-system as n+n+$\alpha$}
he three-body model developed above can be tested on the
$^{6}$He-system, which has been studied as the simplest prototype of a
halo nucleus. The advantage is that the details of the low-energy
two-body interactions are very well known experimentally and the
particles only have high-lying excited states.  The resulting
three-body properties are therefore much less uncertain and related to
the technique rather than to the lack of information about the
subsystems. In this section we shall first study the influence of the
remaining uncertainties in the model, then predict physical properties
of the three-body system and along the way compare with available
data.

\subsection{Interactions and numerical details}
We consider $^6$He as two neutrons and an inert $^4$He-core. The
two-body interactions should in principle only reproduce the
low-energy scattering data which exclusively influence the
computations of spatially extended halo systems. Except for very
accurately needed details it is even quantitatively sufficient to
reproduce the scattering lengths and the effective ranges of the
appropriate partial waves. This initial conjecture \cite{joh90} is now
confirmed and details of the short-range behavior of the two-body
interactions are not needed
\cite{fed94a,fed95,cob97,jen97}. We shall therefore
essentially always maintain the same radial shapes of the
interactions.

The neutron-neutron interaction reproduces the low-energy properties
of free nucleon-nucleon scattering. We have tried several
parametrizations, i.e. the simple neutron-neutron $s$-wave potential,
--31 MeV $\exp(-r^2/(1.8{\rm fm})^2)$, from \cite{joh90}, the extension
to other partial waves in \cite{gar96}, the accurately adjusted
nucleon-nucleon potential from \cite{cob97} and previously known
standard potentials as that of \cite{gog70}. The three-body results
can hardly be distinguished from each other and we shall here only
present results with the interaction from \cite{gar96}.

The neutron-$\alpha$ interaction is parametrized to reproduce
accurately the s-, p- and d-phase shifts up to 20 MeV. We use again
gaussians for the radial shape and allow an $\ell$-dependence of
strengths and ranges, i.e.
\begin{eqnarray}\label{e150}
  V_{nc}^{(\ell=0)} = 48.00  \exp(-r^2/2.33^2)  \nonumber \\
  V_{nc}^{(\ell=1)} = -47.40  \exp(-r^2/2.30^2) 
 -25.49 {\bf \ell} \cdot {\bf s}_n \exp(-r^2/1.72^2)  \\
  V_{nc}^{(\ell=2)} = -21.93 \exp(-r^2/2.03^2)
  -25.49 {\bf \ell} \cdot {\bf s}_n \exp(-r^2/1.72^2)  \nonumber \; ,
\end{eqnarray}
where the strengths are in MeV, the lengths are in fm, ${\bf s}_n$ is
the neutron spin and ${\bf \ell}$ is the relative orbital angular
momentum. The repulsive $s$-wave potential corresponds to a scattering
length of --2.13 fm and an effective range of 1.38 fm. The energies and
widths of the $p$-resonances defined as poles of the $S$-matrix are
$E(p_{3/2})=0.77$ MeV, $\Gamma(p_{3/2})=0.64$ MeV, $E(p_{1/2})=1.97$
MeV and $\Gamma(p_{1/2})=5.22$ MeV, respectively.  The phase shifts
from this potential are in Fig. 1 compared with the results obtained
from scattering experiments \cite{bon77,ali85}.

\begin{figure}\label{fig1}
\centerline{\psfig{file=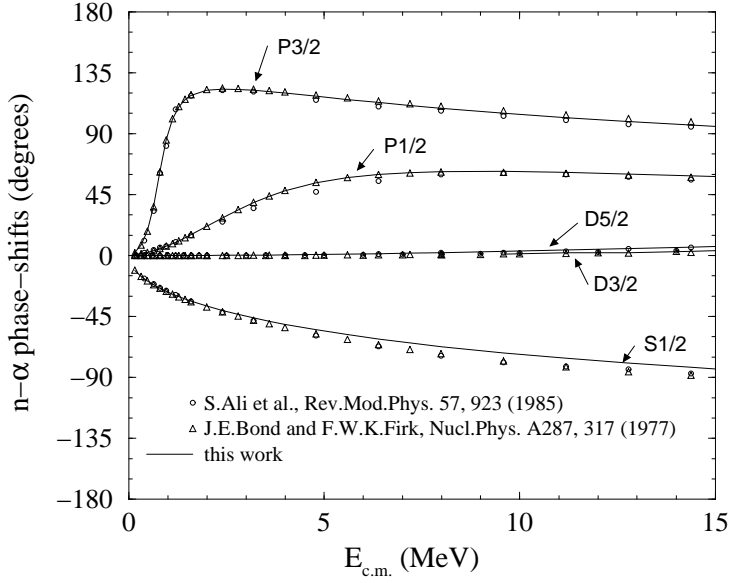,width=10cm,%
bbllx=0cm,bblly=0cm,bburx=19cm,bbury=16cm}}
\caption{ The computed $s$-, $p$- and $d$-wave neutron-$\alpha$
phase shifts (solid curves) compared to the values (triangles and
circles) extracted from scattering experiments
\protect\cite{bon77,ali85}.  The interactions are given in
eq.(\ref{e150}).
}\end{figure}

Other parametrizations are possible even for the same radial shape of
the two-body potential. They differ in the number of two-body bound
states of which the lowest $s$-state is occupied for $^6$He by the
core neutrons and therefore subsequently has to be excluded in the
computation.  For one bound $s$-state the interaction for $\ell=0$ is
\begin{equation}\label{e154}
  V_{nc}^{(\ell=0)} = -75.06 \exp(-r^2/1.53^2) 
\end{equation}
while the $\ell=1,2$ partial waves remain the same as in
eq.(\ref{e150}). The $s$-wave scattering length and the effective
range are the same as for eq.(\ref{e150}) although the potential now
is attractive.

The three-body system computed from these two-body interactions is
underbound by about 500 keV. The required fine tuning is now obtained
by adding a diagonal three-body force $V_3(\rho)$ in
eq.(\ref{e30}). The idea of using the three-body force is to include
effects beyond those accounted for by the two-body interactions. Thus
two-body polarization effects are already included via the effective
two-body interaction. The remaining part must then involve all three
particles simultaneously polarizing each other and therefore only
effective at small $\rho$-values. We therefore use a three-body
interaction only depending on $\rho$ and still allowing for a
dependence of the total angular momentum of the system. We tried both
gaussian and exponential shapes, i.e. $V_3(\rho) = S_{3g} \exp(-\rho^2
/b^2_{3g})$ and $V_3(\rho) = S_{3e} \exp(-\rho /b_{3e})$.

The range of the three-body force is by its definition related to the
hyperradius. For $^6$He, $\rho$=2 fm and 3 fm correspond roughly to
configurations where the neutrons respectively are at the surface of
the $\alpha-$particle and outside the surface by an amount equal to
their own radius. This distance can now be used directly as the range
parameter or defined as the distance where the three-body potential
assumes half of its central value. This means that $b_{3e} = b_{\rho}
/\ln 2$ and $b_{3g} = b_{\rho} /\sqrt{\ln 2}$, where $b_{\rho}=2$ fm or
3 fm for the two different geometric configurations.

The strength of the three-body interaction is finally for $0^+$
adjusted to give the measured two-neutron separation energy 0.97 $\pm$
0.04 MeV of $^6$He.  The different ranges and radial shapes has an
influence on the spatial extension of the three-body system. For
gaussian shapes we obtain root mean square radii of 2.45 fm for both
the attractive $s$-wave potential with one bound state and the repulsive
potential without bound states. The corresponding three-body
interaction parameters are respectively $b_{3g}=2.9$ fm,
$S_{3g}=-$7.55 MeV and $b_{3g}=3.0$ fm, $S_{3g}=-$3.35 MeV.  For
exponential shapes and repulsive $s$-wave potential, we obtain instead
root mean square radii varying almost linearly from 2.61 fm to 2.56 fm
for $S_{3e}=-$3.11 MeV, $b_{3e}=4.3$ fm to $S_{3e}=-$4.77 MeV,
$b_{3e}= 3.0$ fm.  For $2^+$ we could instead fine tune to the well
known resonance of energy 0.820 $\pm$ 0.025 MeV and width 0.113 $\pm$
0.020 MeV \cite{ajz88}. The three-body interaction parameters would
then be J-dependent. For the repulsive potential we obtain
$b_{3g}=2.061$ fm and $S_{3g}=-$31 MeV for gaussian shapes.

In the computations we include all possible $s$-, $p$- and $d$-waves. We use
a hyperspherical basis for each of the Faddeev components with
$K$-values up to about 150. The radial equations are integrated from zero up
to $\rho$-values about 180 fm. Further arguments for these numerical
choices can be found in \cite{cob97a}.

\subsection{Solutions and $S$-matrix poles}
The angular eigenvalues $\lambda_n$ are computed from eq.(\ref{e65})
for total angular momentum and parity $J^{\pi}=0^{\pm}, 1^{\pm},
2^{\pm}$. These eigenvalues are closely related to the effective
potentials in the radial equation eq.(\ref{e30}). Their large-distance
behavior is essential and sometimes decisive as seen in the extreme
case of Efimov states which owe their existence to a sufficiently
negative value of $\lambda$ at very large distance
\cite{fed93,fed94c}. We show in Fig. 2 these spectra for the lowest
spins and parities both computed numerically with the appropriate
basis size and from the analytic expressions for the coupled
$s$-waves.  

\begin{figure}\label{fig2}
\centerline{\psfig{file=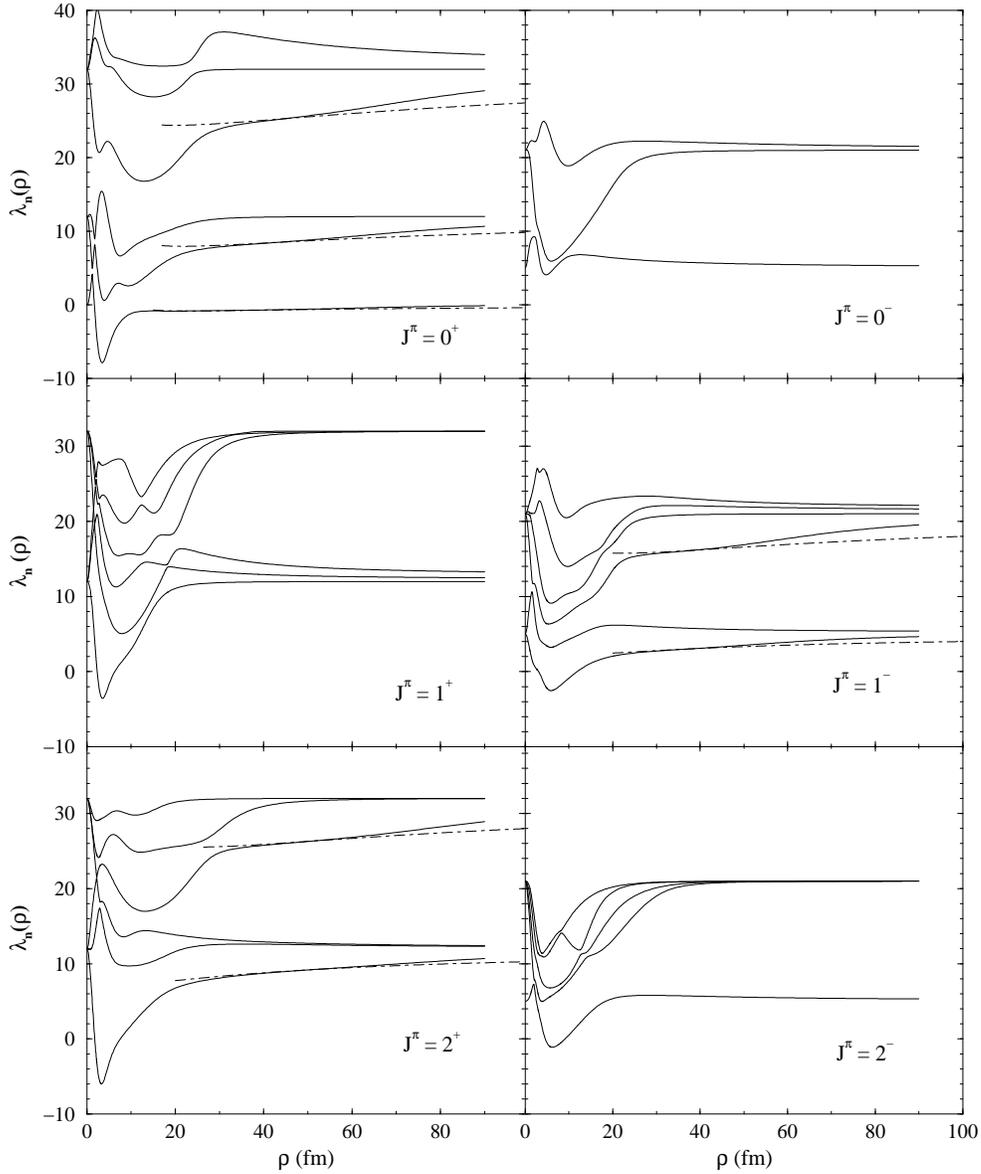,width=13cm,%
bbllx=1.5cm,bblly=3.5cm,bburx=18.5cm,bbury=24.5cm}}
\caption{ The lowest angular eigenvalues $\lambda_n$ for
$^6$He (n+n+$\alpha$) as functions of $\rho$ for angular momentum and
parity $J^{\pi}=0^{\pm}, 1^{\pm}, 2^{\pm}$.  The solid lines are
computed by numerical integration and the dot-dashed lines are the
large-distance asymptotic behavior obtained from eqs.(\ref{e140}) and
(\ref{e141}). The neutron-neutron interaction is from
\protect\cite{gar96} and the neutron-$\alpha$ interaction is from
eq.(\ref{e150}).  Maximum $K$-values up to about 150 are used in the
basis.
}\end{figure}

The asymptotic behavior obtained at large distance is in this case
reached around 40 fm.  This does not mean that all interactions
produce the same results at such a distance. It means that details are
unimportant, but the scattering lengths are still crucial.  A larger
basis would have reproduced the results of the analytical calculations
up to higher $\rho$-values. However, this is not needed, because we
use the asymptotic solutions as soon as they are accurate enough. This
improves both accuracy and speed of the computations. The finite size
of the basis gives a too fast convergence to the hyperspherical
spectrum. Without an independent calculation it can therefore be
difficult to assess the accuracy.

The lowest level for each $J^{\pi}$ usually contribute with the
largest components of the wave function of both the possible bound
state and the low-lying continuum states. For all six cases in Fig. 2
we find pockets which, except for the $0^{+}$ ground state, are
unable to bind the system, but still responsible for several low-lying
$S$-matrix poles as we shall see later. 

The Pauli principle prohibits the core neutrons and the halo neutrons
from occupying the same orbits. This has to be incorporated
explicitly, since the three-body model only deals with particles
without intrinsic degrees of freedom, except for their intrinsic
spins.  For a repulsive $s$-wave potential as described above, no bound
state is present and no overlap has to be excluded. We also
investigate another approximation where we use an attractive
neutron-core potential with one bound $s$-state and the same scattering
length and effective range. The lowest angular eigenvalue must then
bend over and diverge parabolically towards $-\infty$. This
corresponds at large distances to a configuration where a neutron is
bound in the doubly degenerate lowest $s$-state, which is Pauli
forbidden for the halo neutrons. At smaller distances the probability,
or the wave function, must be small, because otherwise a significant
part of the halo wave function would be inside the core, the halo and
core degrees of freedom would not separate and the three-body model
would not be a good approximation. The effective potential at these
small distances is then rather unimportant if the model is valid.
Therefore a good and inexpensive approximation to include the Pauli
principle is simply to omit the lowest diverging angular eigenvalue
$\lambda$ from the computations \cite{gar96}.

\begin{figure}\label{fig3}
\centerline{\psfig{file=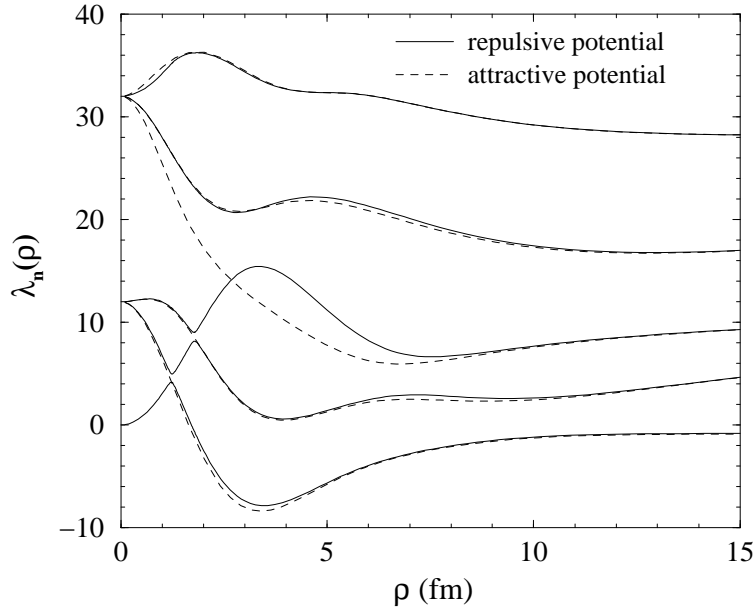,width=10cm,%
bbllx=1cm,bblly=1cm,bburx=19cm,bbury=16cm}}
\caption{ The lowest angular eigenvalues $\lambda_n$ for
$^6$He (n+n+$\alpha$) as functions of $\rho$ for angular momentum and
parity $J^{\pi}=0^{+}$ for the repulsive (solid) and attractive
(dashed) $s$-wave potentials in eqs.(\ref{e150}) and (\ref{e154}).
The neutron-neutron interaction is the same as in Fig. 2. The lowest
diverging level for the attractive potential is omitted corresponding
to our prescription for excluding the Pauli forbidden state.  Maximum
$K$-values up to about 150 are used in the basis.
}\end{figure}

In Fig. 3 is shown the angular eigenvalue spectrum for $J^{\pi}=0^{+}$
for both the repulsive and the attractive $s$-wave potentials in
eqs.(\ref{e150}) and (\ref{e154}). The lowest diverging level for the
attractive potential originating from zero is removed from the figure
as well as from the subsequent computations. The second level for the
attractive potential, originating from 12, is almost identical to the
lowest level from the repulsive potential from about $\rho=1$ fm. The
levels from these two potentials are remarkably similar even at
smaller distances and they are completely identical in the
large-distance asymptotic region. (Note that the figure only shows
results up to 15 fm, where differences still can be seen.) One level
originating from 32 must cross an empty region, and therefore deviate
somewhat from all other levels, until it catches up with one of the
levels from the other potential approaching 12 for large $\rho$.
However, the lowest $\lambda$-value(s) is dominating in the wave
functions of interest here and we therefore should focus on the
corresponding effective potentials.  The differences in these
potentials are small, but in precise computations they must be
compensated in one way or another. Fortunately the means for such fine
tuning is already present as a three-body potential, which is
different for different two-body potentials.

\begin{table}[t]
\renewcommand{\baselinestretch}{0.9}
\caption{The real and imaginary values $(E_r,\Gamma)$ (in MeV) of the
two lowest $S$-matrix poles $E=E_r-i\Gamma/2$ for $^6$He for various
spins and parities.  The interactions used in the upper part of the
table are the same as in Fig. 4. The three-body interaction parameters
are $S_{3g}=-7.55$ MeV, $b_{3g}=2.9$ fm, $S_{3g}=-31$ MeV,
$b_{3g}=2.061$ fm, respectively for the first two and the last two
columns.  The $1^-$ poles in the middle are obtained with the same
interactions except for an exponential shape for the three-body
potential with parameters $S_{3e}=-3.11$ MeV, $b_{3e}=4.3$ fm. In the
lower part of the table the repulsive potential is substituted by the
attractive potential in Eq.(\ref{e154}). The interactions used are
otherwise unchanged, except for the three-body interaction parameters,
where $S_{3g}=-3.35$ MeV, $b_{3g}=3.0$ fm.  The excitation energies
are $E^*=E_r + 0.95$ MeV, $E^*=E_r + 1.54$ MeV for the left and right
hand side of the table.}
\renewcommand{\baselinestretch}{1.5} 
\begin{center}
\begin{tabular}{c|cc|cc||cc|cc||cc|cc}
$J^{\pi}$ & $E_r$ & $\Gamma$ & $E_r$ & $\Gamma$ & $E_r$ & $\Gamma$ &
 $E_r$ & $\Gamma$  \\ 
\hline 
$0^{+}$ & 0.94 & 0.64 & 1.46 & 0.83 & 0.62 & 0.56 & 1.16 & 0.67 \\
$0^{-}$ & 2.07 & 0.74 & - & - & 2.07 & 0.74 & - & - \\
$1^{+}$ & 1.62 & 0.74 & 2.55 & 0.86 & 1.62 & 0.74 & 2.55 & 0.86 \\ 
$1^{-}$ & 1.11 & 0.42 & 1.67 & 0.58 & 0.95 & 0.38 & 1.43 & 0.56 \\
$2^{+}$ & 1.02 & 0.37 & 1.23 & 0.45 & 0.845 & 0.093 & 1.05 & 0.40 \\
$2^{-}$ & 0.90 & 0.34 & 1.82 & 0.57 & 0.90 & 0.34 & 1.82 & 0.57\\
\hline 
$1^{-}$ & 0.96 & 0.38 & 1.44 & 0.54 &      &      &      &     \\
\hline 
$0^{+}$ & 1.02 & 0.59 & 1.48 & 0.75 &      &      &      &     \\
$1^{-}$ & 1.11 & 0.31 & 1.65 & 0.41 &      &      &      &     \\
$2^{+}$ & 1.03 & 0.44 & 1.26 & 0.35 &      &      &      &     \\
\end{tabular} 
\end{center}
\label{tab1}    
\end{table}

\begin{figure}\label{fig4}
\epsfxsize=12cm \centerline{\epsfbox{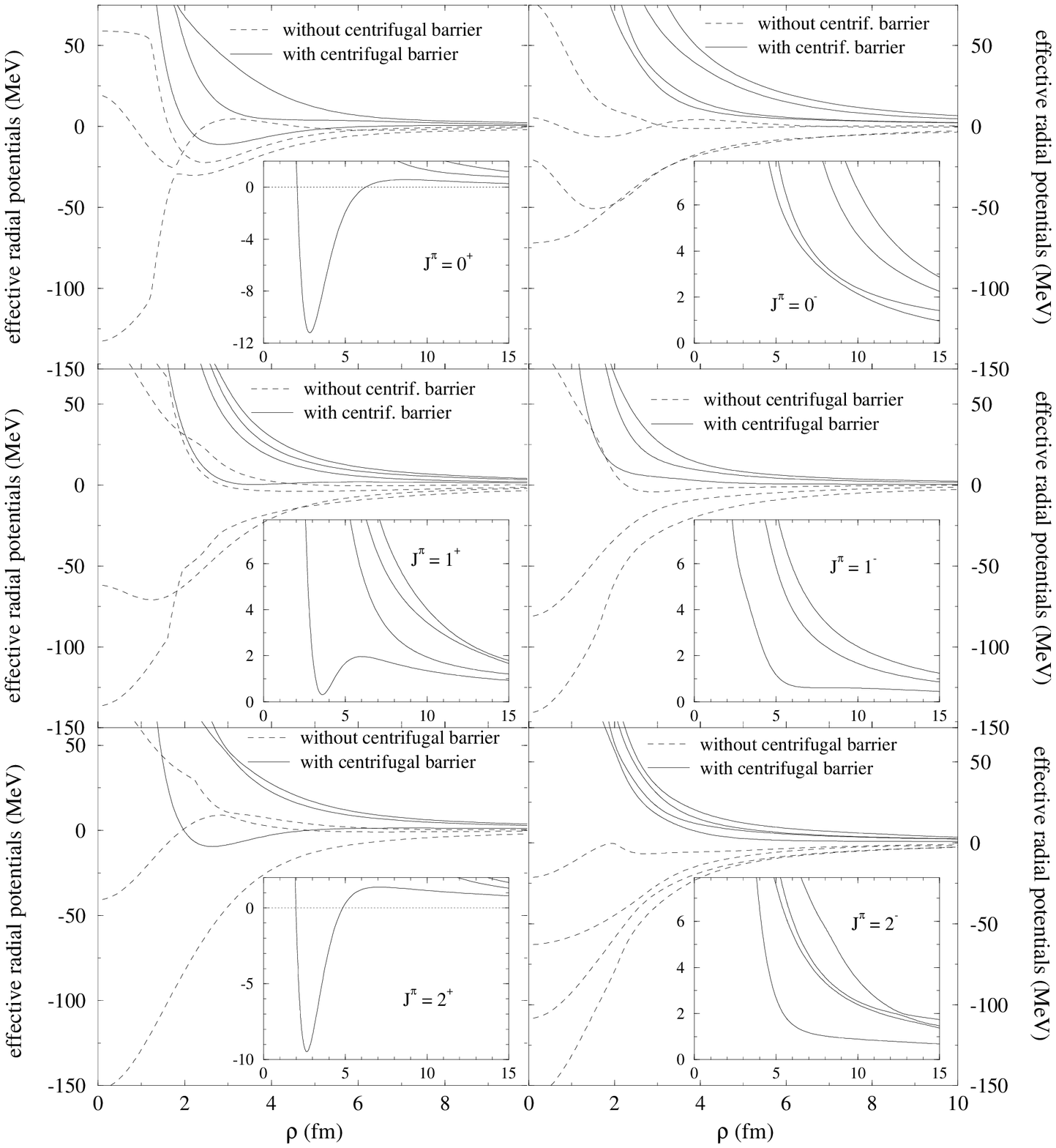}}
\caption{ The total effective diagonal radial potentials
for $^6$He (n+n+$\alpha$) defined in eq.(\ref{e44}) (solid curves) as
functions of $\rho$ corresponding to the three lowest $\lambda$'s for
$0^{\pm}$, $1^{\pm}$ and $2^{\pm}$. The dashed curves are the part
remaining after removal of the generalized centrifugal terms,
i.e. $\hbar^2(K+3/2)(K+5/2)/(2m\rho^2)$, where
$K(K+4)=\lambda(\rho=\infty)$ is the corresponding asymptotic
hyperspherical eigenvalue. The interactions are the same as in Fig. 2
with an additional diagonal three-body interaction, i.e.
$S_{3g}\exp(-\rho^2/b^2_{3g})$ with $S_{3g}=-31$ MeV, $b_{3g}=2.061$
fm, added in all partial waves, except for $J=0^+$ where
$S_{3g}=-7.55$ MeV, $b_{3g}=2.9$ fm. The insets show the details of
the lowest potentials.
}\end{figure}

For the lowest spins and parities we give in Table 1 the lowest
resonance energies and the related widths obtained as $S$-matrix poles
by the complex energy method.  The different three-body forces in
Table 1 can be considered to give the realistic range of the possible
variation in the present model.  In contrast, all previous
computations did not produce three-body resonances in this low-energy
region, except the established $2^+$-resonance
\cite{cso93,aoy95,dan97}. For $0^-$, $1^+$ and $2^-$ we obtain
identical poles, since the two three-body interactions only contribute
at small distances where the effective two-body potentials completely
dominate.  On the other hand, for $0^+$, $1^-$ and $2^+$ we find
differences of up to 0.3 MeV and 0.18 MeV for the position and the
width, respectively.  The systematic shifts of the positions in the
right hand side of the table arise due to the slightly different $2^+$
energy obtained by adjusting the parameters.

The radial shape of the three-body force and the change from the
repulsive to the attractive $s$-wave potential both have an effect on
the lowest $S$-matrix poles. For the exponential three-body force the
numerical values of positions and widths for the two lowest poles are
smaller by about 0.2 MeV and 0.04 MeV, respectively.  The widths are
systematically smaller for the attractive potentials for $0^+$ and
$1^-$ while the $2^+$-poles appear to depend on the individual case.

The widths of these $S$-matrix poles depend rather sensitively on
their energies, which are of the same order as the height of the
corresponding effective radial barriers, see Fig. 4.  For these states
with energies about 1 MeV, any width above 0.4 MeV corresponds to a
smooth structure in the cross sections. Thus, even though the
three-body interactions only amount to a fine tuning of the energies,
the consequences for the presence and subsequent observation of
continuum structures might be substantial.

The low-lying $S$-matrix poles seem to be rather close-lying. By a
sufficiently large additional artificial three-body attraction they
move down towards threshold and become eventually bound states. Their
apparent energies and widths depend rather sensitively on the boundary
condition introduced when the wave functions are matched to the Hankel
functions at a given (large) distance $\rho_{max}$.  In general the
poles move towards the origin until converged with increasing
$\rho_{max}$. Especially the widths are often sensitive to the
matching point. They systematically decrease or sometimes remain
constant with increasing $\rho_{max}$.

For the $2^+$-resonance the sensitivity is very small when
$\rho_{max}$ is larger than 40 fm.  Most of the other poles, however,
require larger $\rho_{max}$ indicating that they are somewhat more
related to the larger distances in hyperradius. This might be a
reminiscence of the Efimov effect, where the bound states are pushed
up into the continuum, but still with a relatively low-lying and dense
energy spectrum. The Efimov effect would arise as the consequence of
very low-lying two-body virtual $s$-states in the neutron-core and the
neutron-neutron subsystems. The actual parameters give energies of
about --200 keV for these virtual $s$-states\footnote{We shall use
negative values corresponding to the energies of the $S$-matrix
poles.}.

The main part of the radial wave function is determined by the angular
momentum dependent effective potential corresponding to the lowest
$\lambda$. We show these potentials for $0^{\pm}$, $1^{\pm}$ and
$2^{\pm}$ in Fig. 4, where we also exhibit the parts remaining after
removal of the generalized centrifugal barrier terms,
i.e. $\hbar^2(K+3/2)(K+5/2)/(2m\rho^2)$, where
$K(K+4)=\lambda(\rho=\infty)$ is the corresponding asymptotic
hyperspherical eigenvalue. As for two-body systems this remaining part
is more revealing than the total potential, which could be repulsive
for all $\rho$ and still produce a resonance provided a sufficiently
strong attractive pocket is present in this ``non-centrifugal'' part
of the potential.

The pocket in the effective radial potential is absent for angular
momentum $0^{-}, 1^{\pm}$ and $2^{-}$. The pocket for $2^+$, which
definitely produces a narrow resonance at about 1 MeV, is slightly
less pronounced than for $0^+$, where a bound state at about 1 MeV is
present.  All the lowest effective potentials are attractive without
the ``centrifugal barrier'' and therefore they could give rise to
resonances.

\begin{figure}\label{fig5}
\centerline{\psfig{file=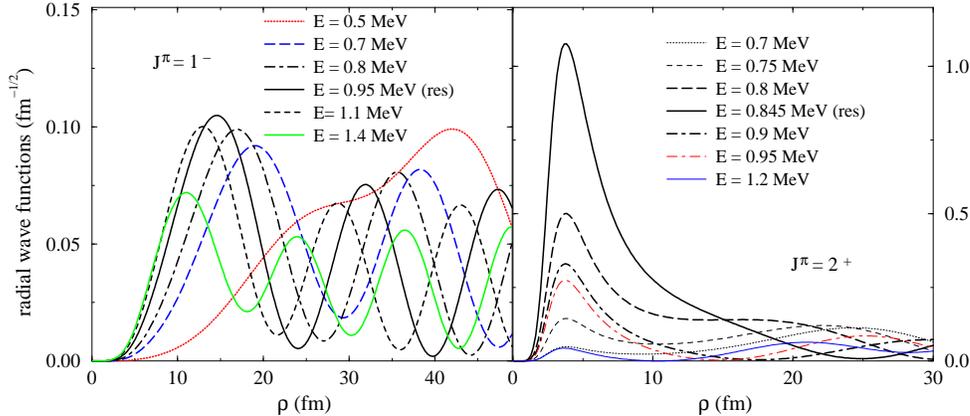,width=13cm,%
bbllx=1.3cm,bblly=10cm,bburx=19cm,bbury=18cm}}
\caption{ The absolute values of the radial wave functions
for $^6$He (n+n+$\alpha$) after diagonalization as functions of $\rho$
for real energies in intervals around the real parts of the $S$-matrix
poles $E_r = 0.95$ MeV for $1^-$ and $E_r = 0.845$ MeV for $2^+$.
Only the dominating component in the full computation is shown.  The
interactions are the same as in Fig. 4.
}\end{figure}

The observables are related to real values of the energy. We therefore
solve the radial equations on the real axis for energies corresponding
both to the real values of the $S$-matrix pole and to values away from
this pole. We show in Fig. 5 the absolute values of these wave
functions for $1^-$ and $2^+$. The peak at small distance is narrower
and much more pronounced for $2^+$ than for $1^-$. Still for $1^-$, a
substantial amount of strength is present between 5 fm and 20 fm
whenever the energy is within the width of the $S$-matrix
pole. Outside the widths of all the poles the wave functions appear
with very little probability at distances below 15 fm, see the curve
for $1^-$ with the energy 1.4 MeV.

We have $2^+$ as a pronounced resonance and $1^-$ which only shows up
as a much smaller and broader peak in Fig. 5. These wave functions
reflect the effects on the real axis of the properties of the
corresponding complex $S$-matrix poles. For other energies and angular
momenta a similar picture is found. The small distance enhancements
are obtained whenever the energy is within the width from the energy
of an $S$-matrix pole. These poles therefore produce observable
effects.  However, the size of the effects depends on both the
detailed properties of the poles and the precise definition of the
observable. It is also clear from Fig. 5 that contributions from
$S$-matrix poles may continuosly vary from substantial to vanishing
small.

\begin{figure}\label{fig6}
\centerline{\psfig{file=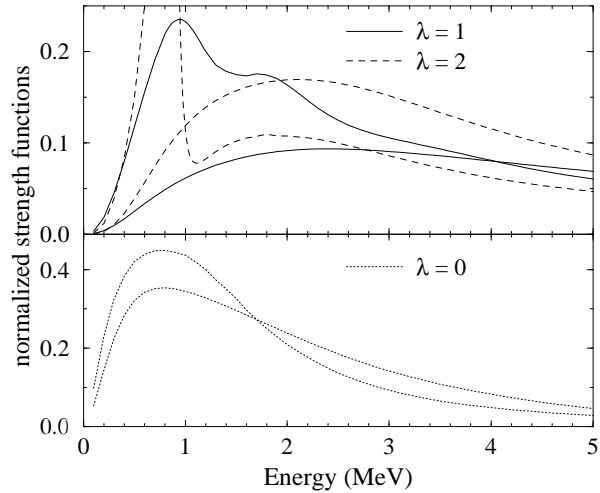,width=8cm,%
bbllx=1.5cm,bblly=1cm,bburx=19cm,bbury=16cm}}
\caption{ The strength functions, $dB_{E\lambda}/dE =
\sum_{n} \left| \langle n J^{\pi} || M(E\lambda) || 0^{+} \rangle
\right|^2$ for $^6$He (n+n+$\alpha$) as functions of energy for
transitions from the ground state to $0^{+}$ (dotted), $1^{-}$ (solid)
and $2^{+}$ (dashed) excited continuum states.  The operator is
$M(E\lambda,\mu) = \rho^2$ and $\sum_{i=1}^3 e Z_i r_i^\lambda
Y_{\lambda \mu}(\hat r_i)$ respectively for $\lambda=0$ and $1,2$. The
units are the corresponding sum rule values $\langle 0^{+}| \rho^4 |
0^{+} \rangle - \langle 0^{+}| \rho^2 | 0^{+} \rangle ^2$ for
$\lambda=0$ and $e^2 Z_\alpha^2 (2\lambda+1) \langle
0^{+}|r_\alpha^{2\lambda} |0^{+} \rangle /(4\pi)$ for $\lambda=1,2$,
where $e Z_\alpha$ is the $^4$He-charge and $r_\alpha$ is the
$^4$He-distance from the $^6$He center of mass.  The interactions are
the same as in Fig. 4. The smooth curves (smaller at small distance)
correspond to plane waves for the continuum states.
}\end{figure}

\subsection{Strength functions and Coulomb cross sections}
The continuum structure can be investigated by electromagnetic and/or
nuclear excitations from the ground state. These transitions are
described by observables such as the multipole strength functions. The
lowest and most important three of these are shown in Fig. 6 as
functions of energy both for plane waves and for the proper continuum
wave functions.  The curves are normalized by their respective sum
rule values and each of them would therefore after integration over
all energies give 1. The strengths below 10 MeV are 97\%, 78\% and
83\% for $\lambda=0,1,2$, respectively and for the corresponding plane
waves we obtain the somewhat smaller values 94\%, 60\% and 81\%
obtained.

The influence of the final state interaction is directly reflected in
deviations from the broader plane wave distributions. In general we
always must have a rise from zero to a maximum and a fall off towards
zero at large energy. Especially pronounced peak structure as observed
for $\lambda=2$ is the signature of a resonance, which in this case
reflects the well known $2^{+}$ state at 0.82 $\pm$ 0.025 MeV of  width
0.113 $\pm$ 0.020 MeV \cite{ajz88}, which in this computation appears
at 0.82 MeV with the width 0.093 MeV. 

For $1^{-}$ a peak and a shoulder appears at about 0.95 MeV and 1.8
MeV. This $1^{-}$ enhancement at low energy arises from the two
overlapping $S$-matrix poles seen in Table 1, see also
\cite{cob97a}. The enhancement almost coincides in energy with the
dominating $2^{+}$-peak and consequently it must be harder to detect
experimentally.  The nuclear $0^{+}$ strength function resembles the
plane wave result more than the higher multipoles reflecting broader
underlying structures where the poles have larger widths if present at
all.

\begin{figure}\label{fig7}
\epsfxsize=13cm \centerline{\epsfbox[42 286 566 510]{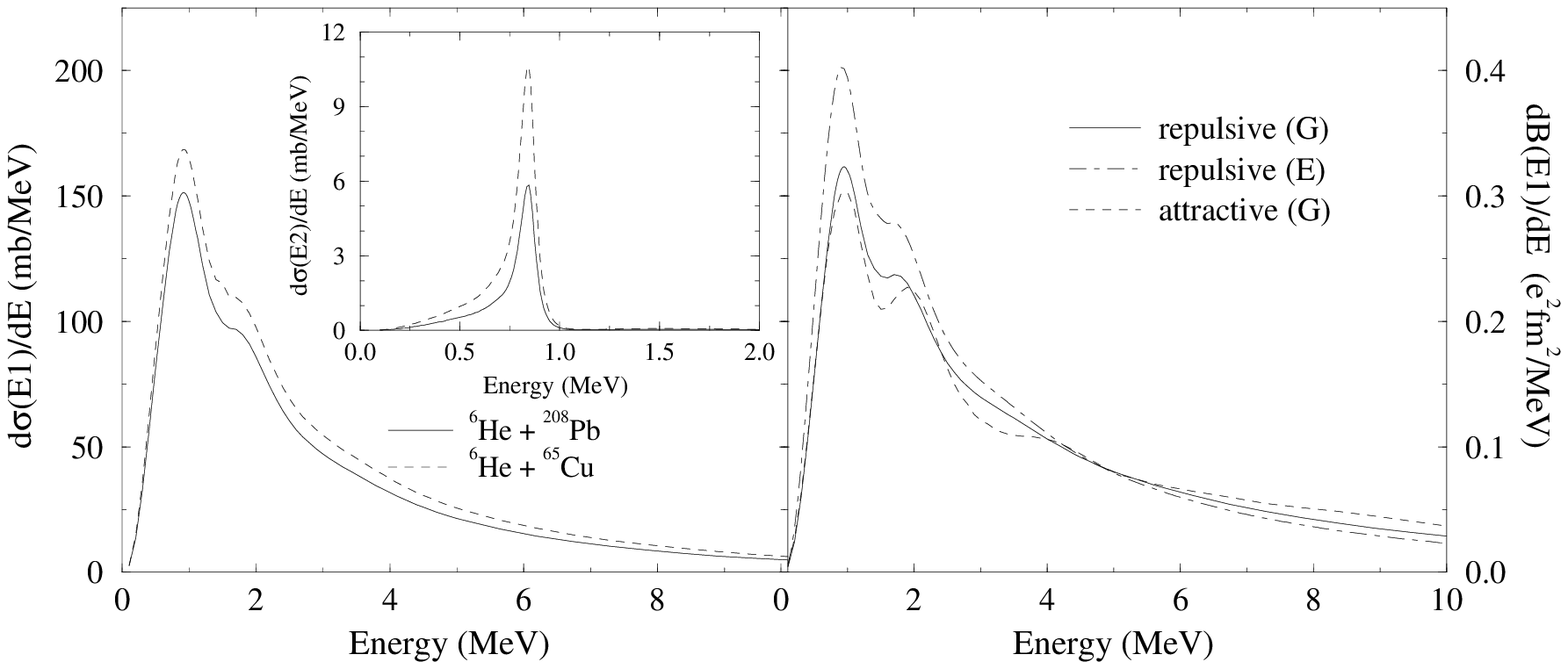}}
\caption{ The dipole and quadrupole (inset) contribution to
the differential Coulomb dissociation cross section of $^{6}$He
(n+n+$\alpha$) at 800 MeV/nucleon as function of the three-body energy
for a Cu- and a Pb-target.  The Cu-results are multiplied by
$(82/29)^2$ to remove the dominating, but trivial overall charge
scaling. The energy of the $^{6}$He-beam is 800 MeV/nucleon.  The
interaction is the same as in Fig. 4. On the right hand side is shown
the strength functions $dB_{E1}/dE$ for the attractive potential in
eq.(\ref{e154}) (dashed) and for the repulsive $s$-wave potential in
eq.(\ref{e150}) (solid and dot-dashed) with gaussian (G) and
exponential (E) three-body interactions. The parameters are given in
section 3.
}\end{figure}

The differential Coulomb dissociation cross section is now computed by
multiplication of strength functions and virtual photon spectra. The
results are shown in Fig. 7 for Pb and Cu targets for both dipole and
quadrupole excitations. As expected the dipole contribution has a
width of about 2 MeV and it is by far dominating in absolute size. The
quadrupole distribution is much smaller, but strongly peaked at the
resonance energy. The target dependence vary with the square of the
target charge as seen from eq.(\ref{e62}).  Both potentials as well as
different three-body interactions give similar, but distinguishable
results as seen in the right hand side of Fig. 7. The major
differences arise from the difference in the ground state structure,
in particular the larger spatial extension found for the exponential
three-body force. With the same ground state wave function almost
identical strength functions would appear.

Previous computations of $1^-$-strength functions reported peaks at
about 2.5 MeV and shoulders at about 6 MeV
\cite{dan93b,cso93,dan97,fun94}. The present $1^-$-strength function
differs substantially with much more strength at low energies
indicating contributions from larger distances.  Unfortunately
corresponding experiments are so far not available for $^{6}$He.

The low-energy enhancement of the dipole strength function move
strength towards energies with larger values of the number of virtual
photons. The total Coulomb dissociation cross section is therefore
larger than that obtained with plane waves in the final state.  It is
also necessarily large compared to analogous cross sections for
ordinary nuclei, again due to a relatively large low-energy
enhancement. This is explained physically as the result of the weakly
bound neutrons easily separated by a small Coulomb disturbance.

The total Coulomb dissociation cross section is simply obtained by
integrating the differential cross section over energy.  The
quadrupole contribution amounts here to about 0.5\% and the total
cross section is 373 mb and 54 mb for the two targets and the beam
energy of 800 MeV/A. This is in agreement with the experimental
extrapolation of \cite{kob90} and the calculated values in
\cite{fer93} while somewhat larger than computed in \cite{suz91}. This
rather favorable comparison supports the three-body model with a
substantial $1^{-}$ low-energy enhancement. However, the enhancement
is not in itself proof of the presence of a low-lying three-body
dipole resonance. Any attraction would produce more strength at low
energies. A resonance needs more than marginal
attraction. Furthermore, the enhancement could be due to relatively
strong underlying two-body structures. In the present case the
$S$-matrix pole at about 1 MeV could indicate a resonance, which however
overlaps the next pole at about 1.5 MeV. This in turn results in the
relatively weak peak in the wave function at small distances in
Figs. 5 and 6.

\begin{figure}\label{fig8}
\epsfxsize=10cm \centerline{\epsfbox[28 28 539 453]{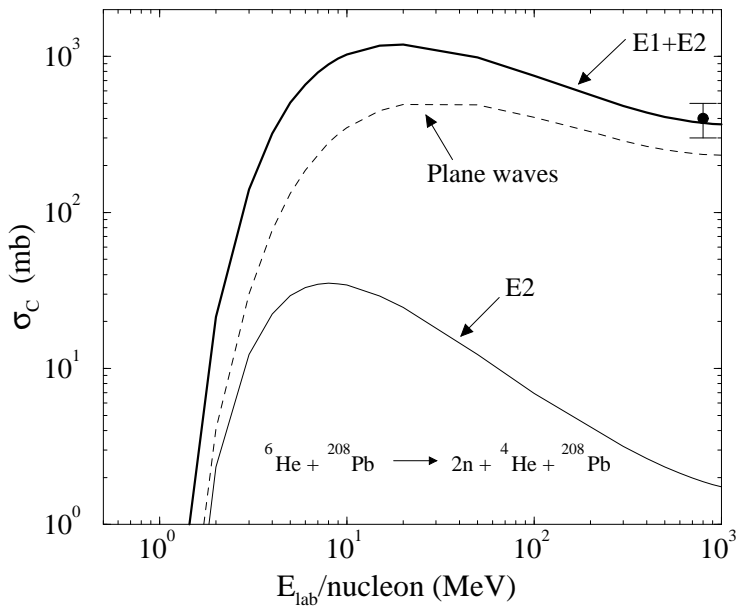}}
\caption{ The total Coulomb dissociation cross section as
function of the laboratory energy for a $^{6}$He-beam colliding with a
$^{208}$Pb-target. Both the quadrupole part and the total cross
sections are shown.  The plane wave results are shown as dashed
curves. The interaction is the same as in Fig. 4. The experimental
point is from \protect\cite{kob90}.
}\end{figure}

The total Coulomb dissociation cross section is shown in Fig. 8 as
function of beam energy. Both dipole and quadrupole contributions are
shown. The experimental point from \cite{kob90} has very large error
bars and therefore not surprisingly in agreement with the
computations.  At high energy we find a slightly decreasing function
with values about a few hundred mb. At energies below about 10 MeV/A
the cross section drops dramatically with decreasing energy. Although
the approximations are dubious at these energies the precise behavior
of this rapid change of the cross section should be sensitive to the
details of the halo structure.

\section{The $^{11}$Li-system as n+n+$^{9}$Li}
The $\alpha$-particle has spin zero and the neutron-$\alpha$ system
has a low-lying $p$-resonance. Consequently $^{6}$He has spin zero and
consists mainly of $p^2$-configurations of the neutron-$\alpha$
relative wave function. For $^{11}$Li the spin and parity is
$\frac{3}{2}^-$ as for the $^{9}$Li-core of the three-body
system. Furthermore the neutron-$^{9}$Li system has apparently a
virtual $s$-states at about --200 keV and somewhat higher-lying
$p$-resonances at about 600 keV \cite{zin97,abr95}. Consequently
$^{11}$Li is expected to consist of dominating $s^2$-configurations of
the relative neutron-$^{9}$Li wave function.  The details about the
two-body subsystems are not accurately known, but already the
possibility of low-lying $s$-states is interesting, since the
conditions for the Efimov effect then nearly are fulfilled. In this
section we shall therefore use the knowledge obtained from the simpler
and better known $^{6}$He and predict the more uncertain properties of
$^{11}$Li.

\subsection{Interactions and numerical details}
The neutron-neutron interaction is the same as used in the previous
computations for $^{6}$He. The neutron-core effective interaction
often assumes zero spin for both $^{9}$Li and $^{11}$Li although the
correct spins are $\frac{3}{2}$ for both nuclei.  The spin-orbit term
${\bf \ell}\cdot{\bf s}_n$ for a neutron in the relative motion around
a spin-zero core is used although the natural generalization, which in
fact has been used previously \cite{gar96,gar97b}, would be of
the form ${\bf \ell}\cdot ({\bf s}_n + {\bf s}_c)$ for finite core
spin $s_c \neq 0$. 

However, in our case of nuclear clusters the Pauli principle must be
treated in one way or another. This can be achieved either by a large
repulsion in forbidden states or by omitting or projecting out the
forbidden configurations. These different approximations assume that
the forbidden states can be identified and preferentially expressed in
terms of the three-body wave functions obtained as solutions for the
corresponding neutron-core potential. Thus an effective two-body
potential is much easier to apply in three-body computations when its
symmetries, quantum numbers and the related eigenfunctions are
expressed by the quantities used in calculations of the nucleonic
motion inside the core. We shall therefore use the mean-field
spin-orbit form ${\bf \ell}\cdot {\bf s}_n$, where the spin of the
core does not enter.

In the present work we shall deal with the Pauli principle in several
ways and then compare the results. The neutrons in the core occupy the
lowest $s_{1/2}$ and $p_{3/2}$-states. The occupied $p_{3/2}$-state is
avoided in the three-body computation by using a sufficiently large
repulsive two-body potential. The neutron-$^{9}$Li $s$-wave potential
is either shallow without any bound states or deeper with one bound
state but the same scattering length and effective range. In the
latter case the forbidden three-body configuration is excluded in the
calculations \cite{gar96}.

Another qualitative difference from the zero core-spin computations is
the two possible couplings of the spins of the neutron and the
core. For the neutron-$^{9}$Li the total spin can then be 1 or 2. In
general we therefore also include a spin-spin potential term to
differentiate between these two spin-couplings for each orbital
angular momentum state. Such spin-splitting terms are most likely
present due to the strong spin dependence of the underlying basic
interaction and consequently hard to ignore.

For finite core spin the interactions corresponding to a shallow
$s$-wave potential are
\begin{eqnarray}\label{e165}
  V_{nc}^{(\ell=0)} = (-7.28 - 0.31 {\bf s}_n \cdot {\bf s}_c) 
 \exp(-r^2/2.55^2)  \nonumber \\
  V_{nc}^{(\ell=1)} = \left( 18.25 + 1.47 {\bf s}_n \cdot {\bf s}_c  
 +55 {\bf \ell} \cdot {\bf s}_n \right) \exp(-r^2/2.55^2)   \; .
\end{eqnarray}
The two $s$-wave scattering lengths and effective ranges are (7.65 fm,
4.53 fm) and (10.88 fm, 4.77 fm) corresponding to virtual $s$-states
at $-0.247$ MeV and $-0.140$ MeV for the total spin of 1 and 2,
respectively.  The energies and widths of the $p_{1/2}$-resonances
defined as a poles of the $S$-matrix are $E(p_{1/2})=0.75$ MeV,
$\Gamma(p_{1/2})= 0.87$ MeV and $E(p_{1/2})=1.60$ MeV and
$\Gamma(p_{1/2})= 3.74$ MeV, respectively for spin 1 and 2. In all
cases the high-lying $p_{3/2}$-resonance is not contributing in the
three-body calculations and the Pauli blocking by the core neutrons
are simulated in this way.

In addition to the two-body potentials a diagonal three-body force
could be introduced for fine tuning as for $^{6}$He. However, the idea
of using the three-body force is to include effects beyond those
accounted for by the two-body interactions and too imprecise and too
little information is available about this neutron-core system. It is
therefore at the moment as reasonable to adjust the two-body
interaction instead of adding another uncertainty at this level.

The choice of interaction parameters is dictated by the knowledge of
$^{11}$Li and the accumulating information about the structure of
$^{10}$Li, i.e. a $p$-resonance at about 0.6 MeV, a low-lying virtual
$s$-state and a small spin splitting of these states \cite{zin97,abr95}.
We obtain a three-body energy of about --300 keV reproducing the
$^{11}$Li-binding energy of 295 $\pm$ 35 keV with the corresponding
root mean square radius of 3.34 fm. Furthermore, the calculated
fragment momentum distributions in $^{11}$Li break-up reactions also
compare rather well with measured values \cite{gar96}. Then the
$^{11}$Li ground state wave function has about 80\% and 20\% of $s^2$
and $p^2$-configurations, respectively.

We shall use this ``realistic'' interaction in the investigation of
the continuum properties of $^{11}$Li.  All possible $s$- and $p$-waves
are included. When the large-distance asymptotic behavior is reached
the solutions are obtained from eqs.(\ref{e76}) and (\ref{e89}). The
radial equations are integrated from zero to about 200 fm.  Further
arguments for these numerical choices can be found in
\cite{cob97a}.

\begin{figure}\label{fig9}
\epsfxsize=13cm \centerline{\epsfbox[48 99 524 694]{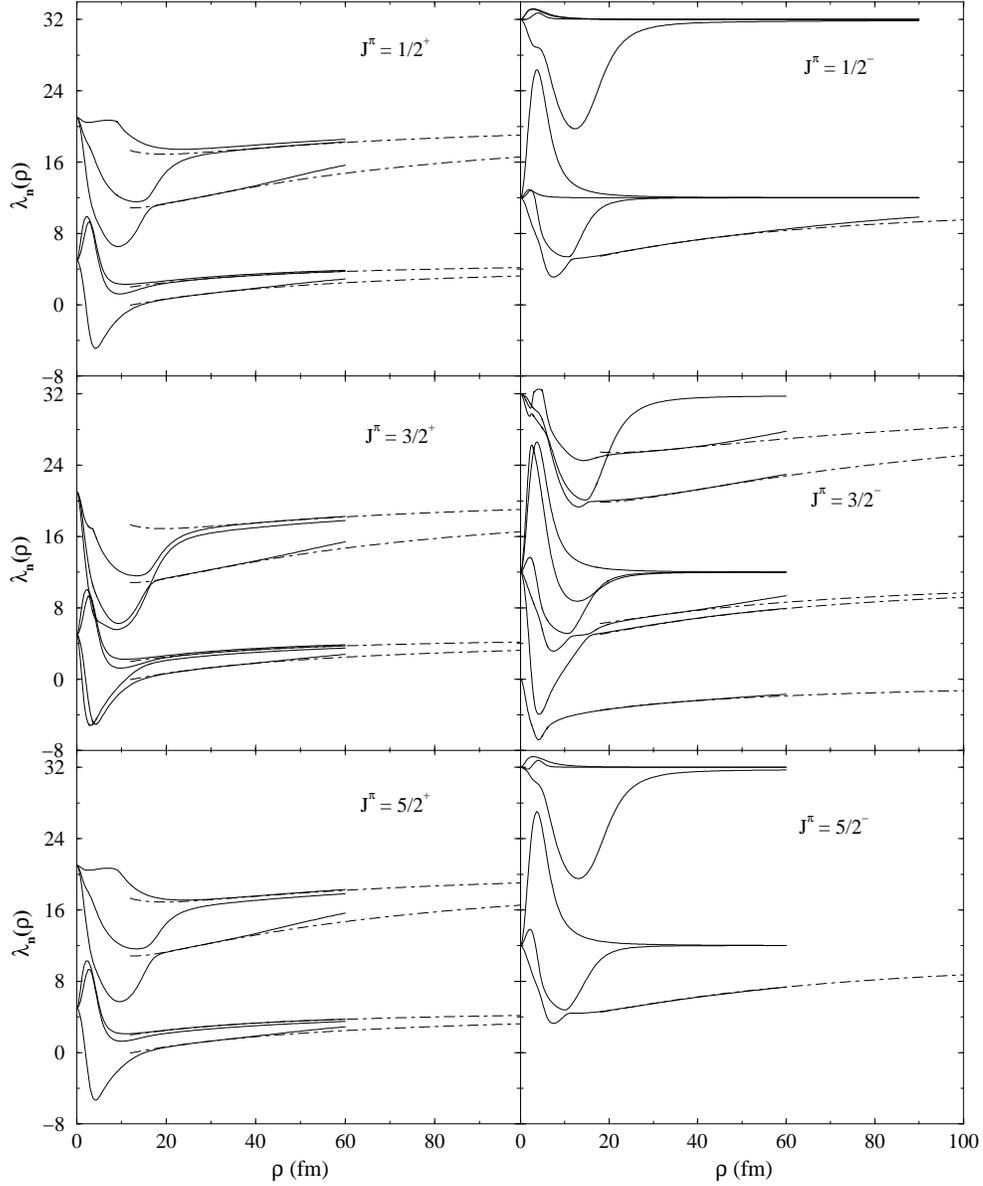}}
\caption{ The lowest angular eigenvalues $\lambda_n$ for
$^{11}$Li (n+n+$^{9}$Li) as functions of $\rho$ for angular momentum
$J^{\pi}=\frac{1}{2}^{\pm}, \frac{3}{2}^{\pm}, \frac{5}{2}^{\pm}$. The
dot-dashed lines are the large distance asymptotic behavior from
eq.(\ref{e90}).  The neutron-neutron interaction is from
\protect\cite{gar96} and the neutron-$^9$Li interaction is given in
eq.(\ref{e165}). Maximum hyperspherical quantum numbers up to about 100
are used.
}\end{figure}

\subsection{Solutions and $S$-matrix poles}

The angular eigenvalues for $\frac{1}{2}^{\pm}$, $\frac{3}{2}^{\pm}$
and $\frac{5}{2}^{\pm}$ are shown in Fig. 9 together with the
asymptotic behavior obtained from the analytical expressions.  For the
ground state with $\frac{3}{2}^{-}$ we find the two lowest levels very
similar to the spectrum for zero core-spin. However, now a series of
additional higher-lying levels appear. They arise from the broken
symmetries due to the finite core spin $\frac{3}{2}$.

The spectra contain almost identical levels for $\frac{1}{2}^{+}$,
$\frac{3}{2}^{+}$ and $\frac{5}{2}^{+}$ as well as for
$\frac{1}{2}^{-}$, $\frac{3}{2}^{-}$ and $\frac{5}{2}^{-}$. A number
of additional levels are furthermore present for $\frac{3}{2}^{\pm}$.
The corresponding degeneracy is due to the weak neutron-core
spin-splitting potential. It can be explained by coupling two neutrons
in $s_{1/2}$ and $p_{1/2}$ neutron-core states to $0^{\pm}$ and
$1^{-}$ which in turn, coupled to the $\frac{3}{2}^{-}$ from the
$^{9}$Li-core, results in sets of nearly degenerate
$\frac{3}{2}^{\pm}$ and $\frac{1}{2}^{+}$, $\frac{3}{2}^{+}$,
$\frac{5}{2}^{+}$-states. The lowest $\frac{1}{2}^{-}$,
$\frac{5}{2}^{-}$-states arise from couplings of higher orbitals.

\begin{table}[t]
\renewcommand{\baselinestretch}{0.9}
\caption{The real and imaginary values $(E_r,\Gamma)$ (in MeV) of the
lowest $S$-matrix poles $E=E_r-i\Gamma/2$ for $^{11}$Li for various
spins and parities $J^{\pi}$. The excitation energy $E^*=E_r + 0.305$
MeV.  The interactions for the upper part of the table are the same as
in Fig. 1. The middle of the table contains the results for a model
with $s_c=0$ and the same average positions of the lowest neutron-core
resonances. i.e. an average energy of the $s_{1/2}$ virtual state and
the $p_{1/2}$-resonance at -0.18 MeV and 1.22 MeV, respectively. The
lower part of the table is for a with one Pauli forbidden
$s$-state. The $s_{1/2}$ virtual state is at -0.18 MeV and the
$p_{1/2}$ neutron-core resonance is at 0.50 MeV. }
\renewcommand{\baselinestretch}{1.5} 
\begin{center}
\begin{tabular}{c|cc|cc|cc|cc|cc|}
$J^{\pi}$ & $E_r$ & $\Gamma$ & $E_r$ & $\Gamma$ & $E_r$ & $\Gamma$ 
& $E_r$ & $\Gamma$ & $E_r$ & $\Gamma$ \\ 
\hline 
$\frac{1}{2}^{-}$ & - & - & - & - & 1.37 & 0.51 & 1.56 & 0.56 & 1.98 & 0.65 \\
$\frac{3}{2}^{-}$ & -0.305 & 0 & 0.89 & 0.43 & 1.41 & 0.56 & 1.60 & 0.61 & 2.03 & 0.68 \\
$\frac{5}{2}^{-}$ & - & - & - & - & 1.36 & 0.49 & 1.60 & 0.68 & 2.01 & 0.72 \\
\hline 
$\frac{1}{2}^{+}$ & 0.65 & 0.35 & - & - & 1.28 & 0.48 & 1.74 & 0.64 & 1.95 & 0.68 \\
$\frac{3}{2}^{+}$ & 0.68 & 0.33 & 0.88 & 0.33 & 1.33 & 0.50 & 1.77 & 0.63 & 2.08 & 0.71 \\
$\frac{5}{2}^{+}$ & 0.68 & 0.37 & - & - & 1.36 & 0.55 & 1.74 & 0.64 & 2.11 & 0.84 \\
\hline 
\hline 
$0^{+}$ & -0.307 & 0 & 1.00 & 0.37 & 1.35 & 0.45 & 1.62 & 0.61 & 1.96 & 0.92 \\
$1^{+}$ & - & - & - & - & 1.40 & 0.59 & 1.59 & 0.63 & 2.02 & 0.81 \\
\hline 
$0^{-}$ & - & - & 0.92 & 0.39 & 1.25 & 0.51 & 1.82 & 0.62 & 2.02 & 0.65 \\
$1^{-}$ & 0.64 & 0.31 & - & - & 1.46 & 0.53 & 1.76 & 0.59 & 2.08 & 0.67 \\
\hline 
\hline 
$0^{+}$ & -0.306 & 0 & 1.00 & 0.31 & 1.40 & 0.41 & 1.64 & 0.56 & 1.99 & 0.88 \\
$1^{+}$ & - & - & - & - & 1.37 & 0.49 & 1.62 & 0.58 & 2.03 & 0.75 \\
\hline 
$0^{-}$ & - & - & 0.88 & 0.30 & 1.29 & 0.43 & 1.76 & 0.53 & 1.96 & 0.69 \\
$1^{-}$ & 0.64 & 0.27 & - & - & 1.50 & 0.39 & 1.82 & 0.60 & 2.00 & 0.71 \\
\end{tabular} 
\end{center}
\label{tab2} 
\end{table}

For the lowest spins and parities we give in Table 2 the lowest
resonance energies and the related widths obtained as $S$-matrix poles
by the complex energy method. We show the results for core-spins zero
with shallow and deep potentials and for the spin $\frac{3}{2}$. The
spin-zero core approximation show a low-lying and relatively narrow
$S$-matrix pole for $1^{-}$, $0^{-}$ and perhaps also for $0^{+}$ while
we find nothing similar for $1^{+}$. Higher-lying and broader poles
are found for all angular momenta and parities.  The same structure is
found for the deep potential. The shallow and deep potentials also
quantitatively give very similar results.

With the correct finite core spin the symmetries are broken. We
recognize the three times nearly degenerate $1^{-}$-pole at about 0.65
MeV with a width of about 0.35 MeV. We also find degenerate $0^{\pm}$
$S$-matrix poles at 0.89 MeV with widths of 0.33 MeV and 0.43
MeV. More discussion about these poles can be found in \cite{cob97a}.

The relatively large number of $S$-matrix poles could be due to the
Efimov effect, which occurs when the scattering lengths are much
larger than the range of the interactions \cite{efi70,fed93}.  With
increasing scattering lengths, the infinitely many poles of the
three-body $S$-matrix move towards the point $E$=0. For very large but
finite scattering lengths a number of poles must already appear close
to zero.  These poles originate from the long distance tail of the
effective potential ($ \propto
\sum_{i=1,3}{a_i}\mu_{jk}^{-1}\rho^{-3}$, where $a_i$ is the
scattering length of the $i$-th subsystem) and they are not sensitive
to the details of the interactions.  Since there are no confining
barriers for these poles, their corresponding widths must be rather
large.

For $^{11}$Li the Efimov condition is almost fulfilled, since
$a_{nn}\mu_{nn}^{-1}+2a_{cn}\mu_{cn}^{-1}\approx$50~fm.  This must
necessarily result in a number of broad $S$-matrix poles near the
$E$=0 point. For $^{6}$He there is a low-lying two-body
$p_{3/2}$-resonance while for $^{11}$Li there is instead an
$s_{1/2}$-virtual state. The latter case is therefore closer to the
Efimov conditions and a larger number of low-lying three-body
$S$-matrix poles could be expected.

An indication of the properties and origin of these $S$-matrix poles
is obtained by matching the radial wave functions at various
decreasing values of $\rho_{max}$. The poles move towards larger
absolute values of the energies for decreasing matching radii. The
imaginary values stay almost constant for the lowest $0^+$ and
$1^-$-states indicating resonance-like structures. Both imaginary
and real values increase for the other poles.

\begin{figure}\label{fig10}
\epsfxsize=13cm \centerline{\epsfbox[34 99 566 694]{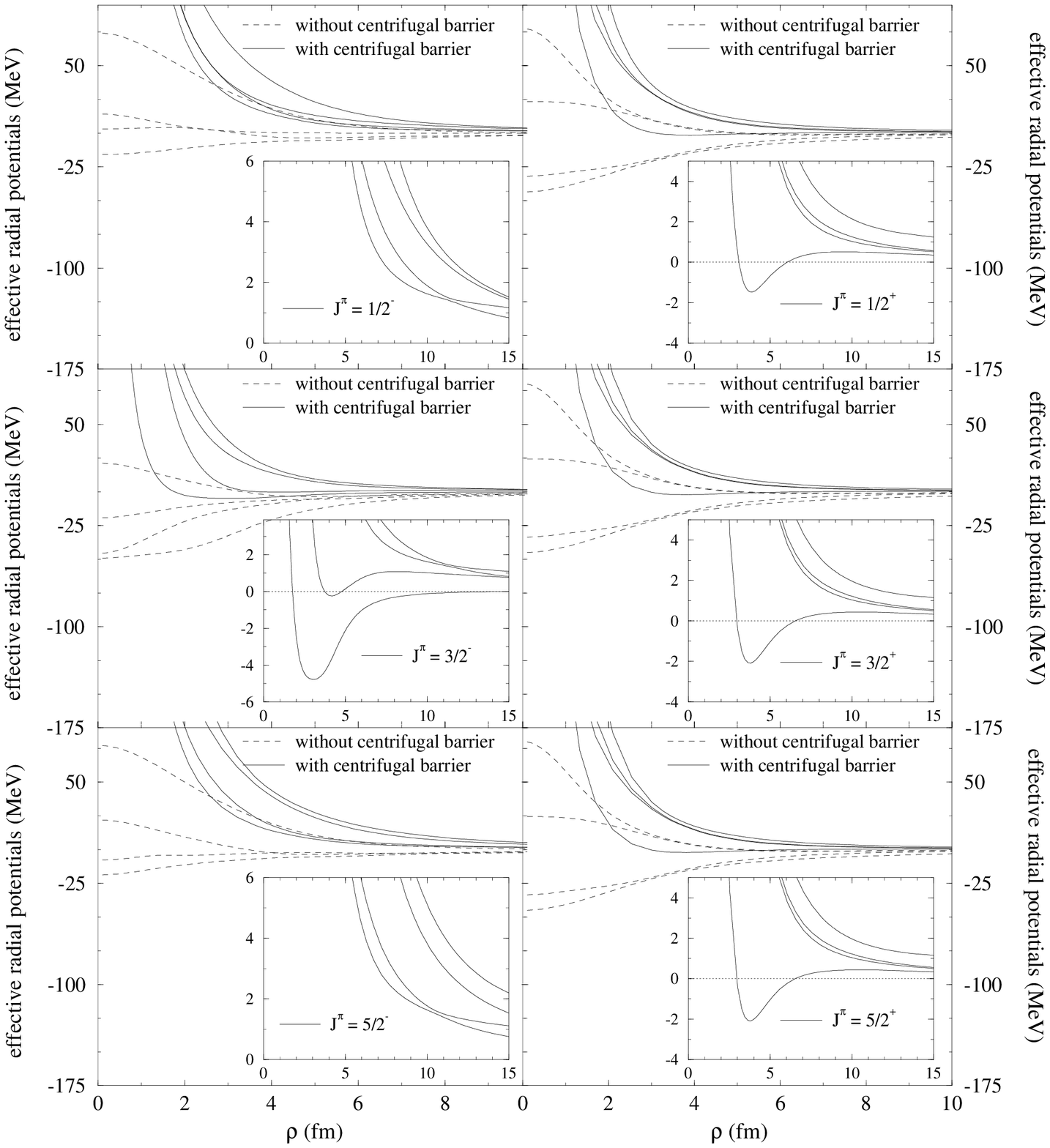}}
\caption{ The total effective diagonal radial potential
(solid curves) as functions of $\rho$ corresponding to the four lowest
$\lambda$'s for $J^{\pi}=\frac{1}{2}^{\pm}, \frac{3}{2}^{\pm},
\frac{5}{2}^{\pm}$ for $^{11}$Li (n+n+$^{9}$Li).  The dashed curves
are the part remaining after removal of the generalized centrifugal
terms, i.e. $\hbar^2(K+3/2)(K+5/2)/(2m\rho^2)$, where
$K(K+4)=\lambda(\rho=\infty)$ is the corresponding asymptotic
hyperspherical eigenvalue. The interactions are as in Fig. 9.  The
insets show the details of the lowest potentials.
}\end{figure}

The lowest effective adiabatic potentials determine the radial wave
function and the energy of possible bound states. They are shown in
Fig. 10 for various spins and parities. In all cases we find
attractive potentials around 50 MeV deep after removal of the
repulsive centrifugal barriers. Resonances are therefore possible in
all these channels. With the centrifugal barriers all the potentials
are still attractive except those corresponding to $\frac{1}{2}^{-}$
and $\frac{5}{2}^{-}$. The ground state of $J^{\pi}=\frac{3}{2}^{-}$
exhibits the largest attractive pocket and no barrier for the lowest
adiabatic potential and a barrier height of 1.7 MeV at about 7 fm for
the second potential. The $1^{-}$ excited states all have attractive
pockets as well as repulsive barriers of about 0.6-0.9 MeV for $\rho$
between 10 and 15 fm.

Compared to $0^+$ of $^{6}$He we have now a less attractive but
broader $\frac{3}{2}^{-}$-potential corresponding to the ground state
quantum numbers, see Fig. 4. The $1^-$ excited states for $^{6}$He
have no attractive pocket while it is substantial for $^{11}$Li in
agreement with the calculated low-lying $S$-matrix poles, which appear
around the barrier height.

\begin{figure}\label{fig11}
\epsfxsize=9cm \centerline{\epsfbox[42 30 524 439]{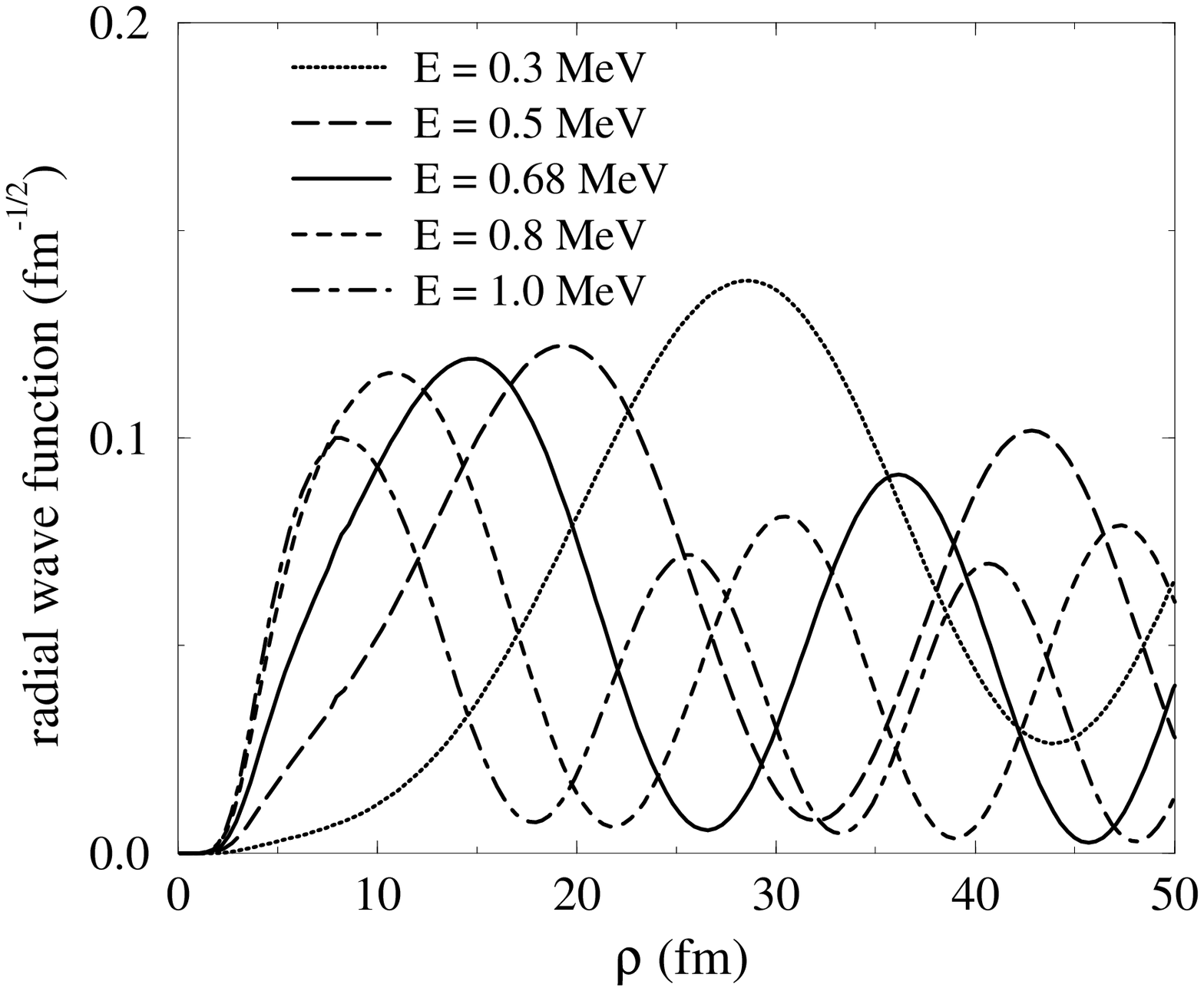}}
\caption{ The absolute values of the radial wave
functions for $^{11}$Li (n+n+$^{9}$Li) as functions of $\rho$ for real
energies in an interval around the real part of the $S$-matrix pole
$E_r = 0.68$ MeV for $\frac{3}{2}^+$. The interactions are the same as
in Fig. 9.
}\end{figure}

The wave functions corresponding to real energies around the real part
of the pole energy 0.68 MeV are shown in Fig. 11. For energies below
0.68 MeV and outside its width the peak moves to larger distances, but
remains of comparable size. For the energy 0.8 MeV we find a similar
peak at a slightly smaller distance. This peak can be viewed as the
combined effect on the real axis of the two overlapping pole
structures at 0.68 MeV and 0.88 MeV, see Table 2. For 1.0 MeV,
respectively within and outside the widths of the poles at 0.88 MeV
and 1.33 MeV, the peak has decreased and moved to a smaller distance.
None of all these peaks are pronounced in comparison with the next
peaks of the same wave function. Thus strong $1^-$-resonance
structures are not obtained.

\begin{figure}\label{fig12}
\epsfxsize=13cm \centerline{\epsfbox[48 283 538 510]{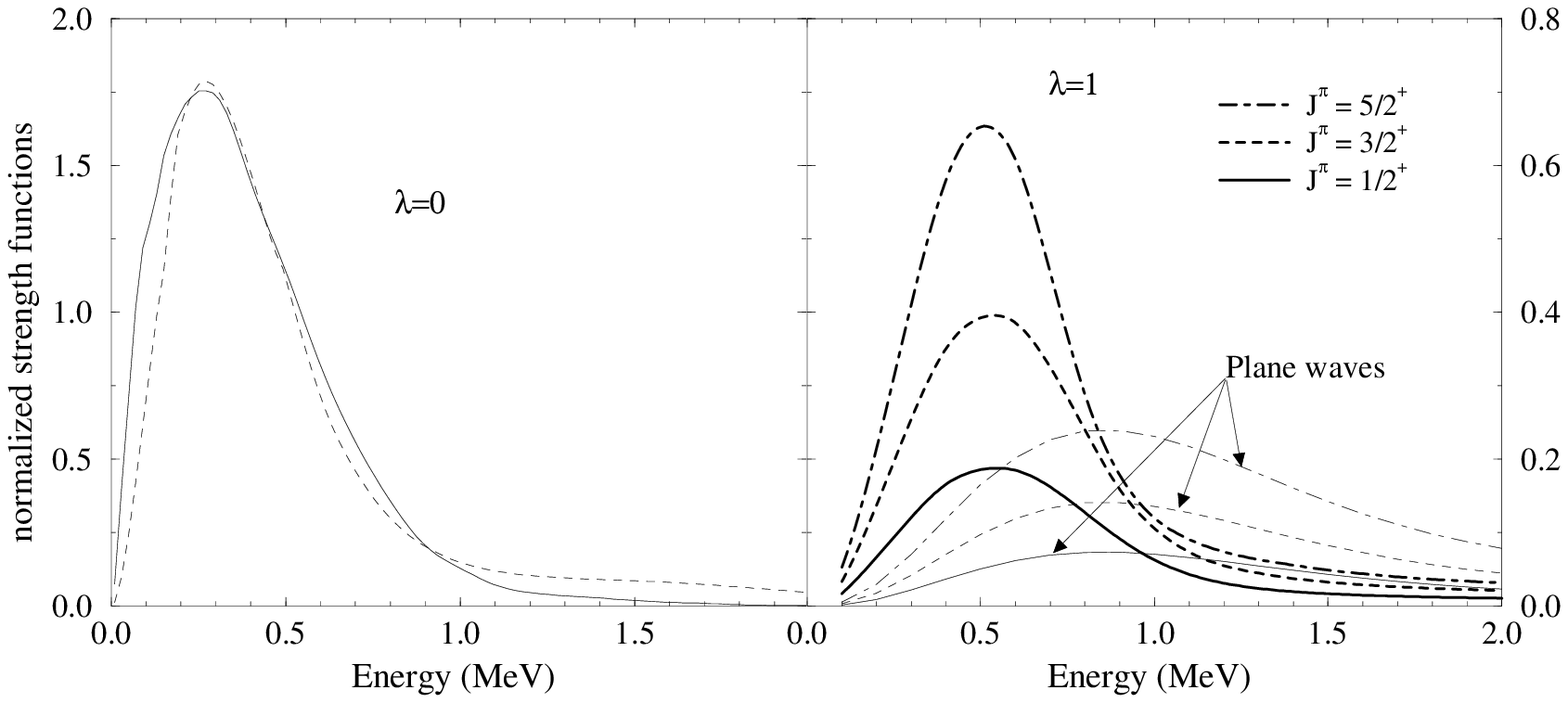}}
\caption{ The strength functions $dB_{E\lambda}/dE
\propto \sum_{n} \left| \langle n J^{\pi} || M(E\lambda) ||
J_0^{\pi_0} \rangle \right|^2$, $M(E\lambda,\mu)$ = $\sum_{i=1}^3 eZ_i
r_i^\lambda Y_{\lambda \mu}(\hat r_i)$, for $^{11}$Li (n+n+$^{9}$Li)
as function of energy for transitions from the ground state via $0^+$
to $\frac{3}{2}^{-}$ (left hand side), via $1^-$ (right hand side) to
$\frac{1}{2}^{+}$, $\frac{3}{2}^{+}$ and $\frac{5}{2}^{+}$ excited
continuum states.  The smooth curves (smaller at small distance)
correspond to plane waves for the continuum states. The curves are
normalized to the corresponding sum rule values given in Fig. 6.  The
interactions are the same as in Fig. 9.
}\end{figure}

\subsection{Strength functions and Coulomb cross sections}
The dominating dipole term in electromagnetic excitations can excite
the ground state to continuum states of $J^{\pi}= \frac{1}{2}^{+},
\frac{3}{2}^{+}, \frac{5}{2}^{+}$ while the nuclear monopole
excitation only produce $\frac{3}{2}^{-}$-states. The corresponding
calculated strength functions are shown in Fig. 12 together with the
results obtained by using plane waves for the continuum wave
functions. The monopole strength resembles the results of the plane
wave computation in agreement with the lack of low-lying $S$-matrix
poles below 0.8 MeV, see Table 2.  The dipole strengths are almost
proportional to the statistical weights of $(2J+1)$ and all of them
are enhanced significantly above the plane wave results at low
energies. This enhancement overlaps with the position of the lowest
$1^-$-poles in Table 2.

\begin{figure}\label{fig13}
\epsfxsize=9cm \centerline{\epsfbox[48 31 538 442]{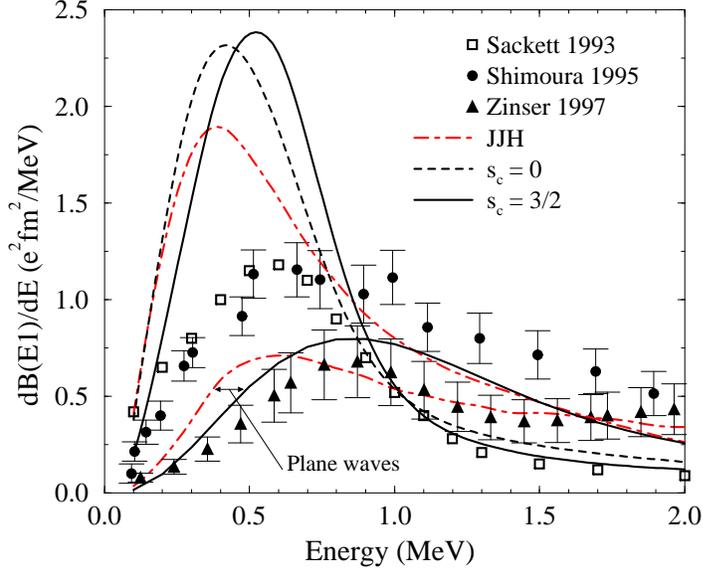}}
\caption{ The strength functions $dB_{E\lambda}/dE =
\frac{1}{2J_0+1}\sum_{nJ^{\pi}} \left| \langle n J^{\pi} ||
M(E\lambda) || J_0^{\pi_0} \rangle \right|^2$, $M(E\lambda,\mu) =
\sum_{i=1}^3 eZ_i r_i^\lambda Y_{\lambda \mu}(\hat r_i)$, for
$\lambda=1$ for $^{11}$Li (n+n+$^{9}$Li) as function of energy.  The
smooth curves correspond to plane waves for the continuum states.  The
interactions are the same as in Table 2 for $s_c=0$ and Fig. 9 for
$s_c=\frac{3}{2}^-$ except for the JJH curve obtained with the
potential in \cite{joh90}. The experimental points are from
\cite{zin97,sac93,shi95}. The arbitrary units in \cite{shi95} are
normalized to our sum rule value, while the absolute data from
\cite{zin97,sac93} are left unchanged.
}\end{figure}

The total dipole strength function, where the contributions from all
$J^{\pi}$ in the continuum are added, is in Fig. 13 compared to the
zero core approximation and the three available measured
distributions. The plane wave result are the same for zero and finite
core spin, because the ground state essentially is unchanged. The
interaction with zero core-spin gives a distribution shifted about 100
keV towards lower energy compared to the result for the realistic full
computation. A lower and broader peak is obtained for the potential
from \cite{joh90} where the $p^2$-content of the three-body wave
function is very small.  The low-lying $1^{-}$-poles around 0.65 MeV
enhance the strength functions at low energies compared to the plane
wave results.

The computed strength functions substantially exceed most of the data
points \cite{zin97,sac93,shi95} in the peak region around 0.55
MeV. (Note however that the data in \cite{zin97,sac93} contain much
less total strength.)  A reduction could be achieved with higher
energy and larger width of the $S$-matrix pole, but this would
probably only be provided by a potential with much too small
$p^2$-content in the three-body wave function. It is also a curious
fact that other models provide $1^-$ strength functions substantially
closer although not in complete agreement with the data
\cite{dan94,esb92,sag96}.  This is of course related to the lack of
low-energy $1^-$-resonances or $S$-matrix poles in these computations.

In this context it is worth pointing out that it is difficult to find
the most appropriate comparison with the different experimental
results in the figure. The normalization must be properly chosen and
the theoretical results must be folded with the distributions (unknown
to us) related to the equipment used in the different experiments.

\begin{figure}\label{fig14}
\epsfxsize=9cm \centerline{\epsfbox[28 28 538 442]{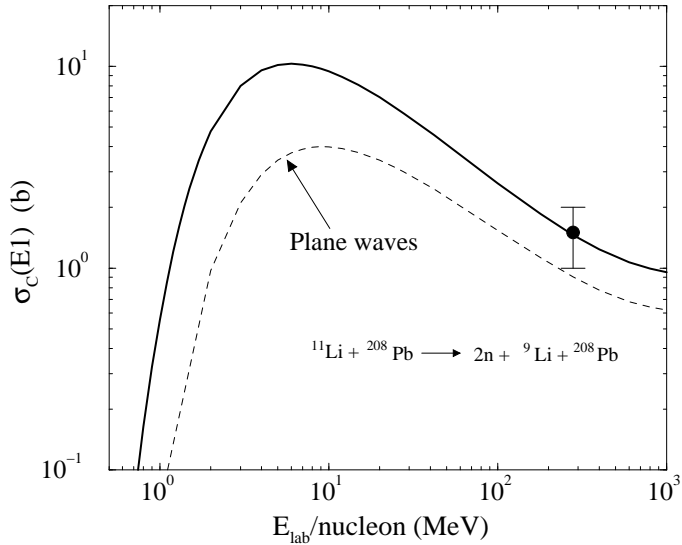}}
\caption{ The dipole approximation (solid curve) to the
Coulomb dissociation cross section of $^{11}$Li as function of
laboratory energy per nucleon for a Pb-target. The plane wave results
are the dashed curve.  The interactions are the same as in
Fig. 9. The data point is from \cite{zin97}.
}\end{figure}

The strength functions rise from zero to a maximum and then fall off
towards zero at large energy whereas the virtual photon spectra
decrease monotonically with excitation energy \cite{ber88}. The
low-energy enhancement necessarily implies a larger Coulomb cross
section, since the dipole is the dominating dissociation mode. We show
in Fig. 14 computed total cross sections as function of beam energy
for a $^{11}$Li projectile on a lead target. The only available data
point with relatively large error bars is in agreement with the
computation \cite{zin97}, but substantially larger than the plane wave
result.  For 280 MeV/A we find 1458 mb and 1501 mb, respectively for
the potentials with zero core-spin and with a small $p^2$-content of
the wave function.

The cross section decreases at large beam energy due to the similar
decrease of the virtual photon spectrum. At small energy, where the
approximations are invalid, we find an increasing cross section due to
the low-energy cut-off of the strength function by the virtual photon
spectrum. At the maximum around 5 MeV per nucleon this and other
reactions are expected to be sensitive to the details of the neutron
halo structure.

\section{Summary and conclusions}

Borromean halo systems are almost by definition weakly bound and
excited states are usually entirely absent. The continuum is therefore
easy to excite and unavoidable in descriptions of essentially all
reactions involving such particles. The spatial extention of the bound
state and the small binding energy require rather accurate treatment
of the large distances. A suitable method was recently developed and
applied to compute the ground state of halo systems. This method is in
the present paper extended to apply for the low-energy part of the
three-body continuum.

The Faddeev equations in coordinate space are solved in two
steps. First the discrete spectrum of the angular part is computed and
used as a complete basis set. Then the coupled set of radial equations
is solved with the appropriate continuum wave boundary conditions. The
angular equations for large distances outside the short ranges of the
potentials are especially simple. However, they are essential and
therefore treated carefully.

The three Faddeev components are very useful in the detailed
description of the particle correlations. We show that only the three
$s$-waves (one in each component) for a given total orbital angular
momentum couple at large distances. All other partial waves are
decoupled. We therefore solve these much smaller and simpler sets of
coupled and uncoupled angular equations at large distances. Some of
these asymptotic solutions are obtained analytically and others as
solutions to trancendental analytical equations.  As a necessary
intermediate result we derive a convenient expression for the
transformation of angular functions between two different Jacobi
coordinate systems.

Systems with two identical neutrons and a core of finite spin are
specifically treated. The continuum spectra of the two Borromean
halo nuclei $^{6}$He (n+n+$\alpha$) and $^{11}$Li (n+n+$^{9}$Li) are
investigated numerically in some detail. Two-body interactions with
and without bound states, but reproducing the observed low-energy
scattering data are used for $^{6}$He. In addition three-body
interactions with several radial shapes are added to obtain the
measured binding energy. For $^{11}$Li no three-body interaction is
used, since the two-body interaction is unknown and therefore directly
parametrized to reproduce anticipated two-body resonances and
virtual states in addition to the momentum distributions in
fragmentation reactions.

The antisymmetry between the neutrons in the halo and in the core is
accounted for in two ways. First by using a repulsive or a shallow
neutron-core potential without bound states. Second by omitting the
lowest adiabatic potential arising from a more attractive neutron-core
potential with one bound state from the set of radial equations. The
results are compared. 

The adiabatic potentials are decisive for the radial solutions.  The
lowest potential in each channel is attractive when the corresponding
generalized centrifugal barrier is removed. All channels are therefore
potentially able to support resonance-like structures. The pocket in
the adiabatic radial potential is absent for $^6$He for angular
momentum $0^{-}, 1^{\pm}$ and $2^{-}$ and well developed for both
$0^+$ and $2^+$. For $^{11}$Li the adiabatic potentials for
$1^-$-excitations all have attractive pockets and effective barriers
of $0.75 \pm 0.15$ MeV. Also the $0^+$-channel has a well developed
attractive pocket, but no barrier, for the lowest potential supporting
the ground state. The second potential for $0^+$-excitations is also
attractive with a barrier of about 1.7 MeV.

We calculate the $S$-matrix poles by the complex energy method.  The
lowest of these poles appear slightly above the barriers and their
widths are consequently relatively large and rather sensitive to fine
tuning of the interactions. One exception is the known narrow
$2^+$-resonance in $^6$He which is reproduced in the calculation. The
narrowest low lying poles for $^6$He appear for $1^-$ and $2^-$ at
about 1 MeV with widths of 0.3-0.4 MeV. For $^{11}$Li they appear for
$1^-$ and $0^{\pm}$ respectively at about 0.65 MeV and 0.9 MeV with
widths of about 0.35 MeV.  The unusually many low-lying $S$-matrix
poles could indicate that the Efimov limit is fairly close.

We computed the electric excitations from ground to continuum
states. The strength functions are rather strongly enhanced at low
energies due to the low-lying $S$-matrix poles. The functions
extracted from measurements for $^{11}$Li are apparently significantly
smaller than our computations. On the other hand the same experimental
information agrees with the computed Coulomb cross section. A proper
consistent comparison is still lacking.  Also for $^6$He we obtain
enhanced dipole strength functions at low energies. Here the
experimental information is not available, but the observed
$1^{-}$-resonance is reproduced almost within the experimental
uncertainties. We have not attempted to reproduce this somewhat
controversial resonance more precisely.

In conclusion, we have developed a method to solve the three-body
problem for short-range potentials. The method treats with special
care the large distances which are essential for the spatially
extended halo systems. We investigate the continuum spectra for the
two halo nuclei $^{6}$He and $^{11}$Li and find a number of low-lying
$S$-matrix poles. Strength functions are computed and compared with
other calculations and available experimental data. Various
disagreements are pointed out and several controversial features are
exhibited.

\paragraph*{\bf Acknowledgments.} We thank E. Garrido and E. Nielsen for 
help and continuos discussions. One of us (A.C.) acknowledges the
support from the European Union through the Human Capital and Mobility
program contract nr. ERBCHBGCT930320.

\bigskip
\noindent
{\bf Appendix A \\ Rotations between different sets of Jacobi
coordinates} \setcounter{equation}{0} \\
\noindent
We want to ``rotate'' the wave function from one set of Jacobi
coordinates to another set as defined in eq.(\ref{e75}). Only the
leading order in an expansion in $1/\rho$ is needed. We must then
first express $({\bf x}_j, {\bf y}_j)$ in terms of $({\bf x}_i, {\bf
y}_i)$. The six-dimensional transformation is \cite{ray70,vin90}
\begin{eqnarray} \label{a1}
{\bf x}_j = - {\bf x}_i \cos\varphi_k + {\bf y}_i \sin\varphi_k \; \; ,
\; \; {\bf y}_j = - {\bf x}_i \sin\varphi_k - {\bf y}_i \cos\varphi_k
\end{eqnarray}
where $\varphi_k$ is defined in eq.(\ref{e83}). Defining the angle
$\gamma$ between ${\bf x}_i$ and ${\bf y}_i$ by
\begin{equation}
   \cos\gamma \equiv {{\bf x}_i \cdot {\bf y}_i \over x_i y_i} \; ,
\end{equation}
we have the relation between the hyperangles $\alpha_i$ and $\alpha_j$
related to the two coordinates
\begin{equation} \label{a3}
   \sin^2\alpha_j = \cos^2\varphi_k \sin^2\alpha_i + 
   \sin^2\varphi_k \cos^2\alpha_i
   +2 \cos\varphi_k \sin\varphi_k \sin\alpha_i \cos\alpha_i \cos\gamma \;.
\end{equation}

We now expand the following function, related to the function in
eq.(\ref{e75}), of $\alpha_j$ in terms of Legendre polynomials
$P_\ell(\cos\gamma)$:
\begin{equation} \label{a7}
\frac{\phi_{n \ell_x^\prime \ell_y^\prime L s_x^\prime S}^{(j) }(\alpha_j)}
{\sin(2\alpha_j)  \sin^{\ell_x'}\alpha_j \cos^{\ell_y'} \alpha_j} =
\sum_{\ell}  A_{\ell}^{\ell_x'\ell_y'L}(\alpha_i)P_{\ell}(\cos\gamma) \;, 
\end{equation}
\begin{equation} \label{a9}
   A_{\ell}^{\ell_x'\ell_y'L}(\alpha_i) \equiv
   {2\ell+1 \over 2}\int d\cos\gamma 
   {\phi_{n \ell_x^\prime \ell_y^\prime L s_x^\prime S}^{(j)} (\alpha_j)
   \over \sin(2\alpha_j) \sin^{\ell_x'}\alpha_j \cos^{\ell_y'}\alpha_j}
   P_{\ell}(\cos\gamma) \;,
\end{equation}
where $\alpha_j$ is the function of $\alpha_i$ and $\gamma$ defined through
eq.(A\ref{a3}). Then changing the integration variable from $\cos\gamma$
to $\alpha_j$, i.e.
\begin{equation} \label{a11}
d\cos\gamma= {\sin(2\alpha_j) \over \sin (2\alpha_i)} 
{2 \over \sin(2\varphi_k)} d\alpha_j \;,
\end{equation} 
we can rewrite eq.(A\ref{a9}) as
\begin{eqnarray} \label{a13}
A_{\ell}^{\ell_x'\ell_y'L}(\alpha_i) =  {2\ell+1 \over \sin(2\varphi_k)
\sin(2\alpha_i)} 
\int_{|\varphi_k-\alpha_i|}^{\pi/2-|\pi/2-\varphi_k-\alpha_i|} d\alpha_j
\nonumber \\ \times
{\phi_{n \ell_x^\prime \ell_y^\prime L s_x^\prime S}^{(j)} (\alpha_j)
 \over \sin^{\ell_x'}\alpha_j \cos^{\ell_y'}\alpha_j}
P_{\ell}(\cos\gamma(\alpha_i,\alpha_j))
\end{eqnarray}

Next we use the following identities
\begin{equation}\label{a19} 
P_{\ell}(\cos\gamma) = \frac{4\pi}{2\ell + 1}
Y_{\ell \ell}^{00}(\Omega_{x_i},\Omega_{y_i})
\end{equation}
 
\begin{eqnarray} \label{a17}
N_{0\ell_x'\ell_y'} \sin^{\ell_x'}\alpha_j \cos^{\ell_y'}\alpha_j
Y_{\ell_{x}^\prime \ell_{y}^\prime}^{LM_L}(\Omega_{x_j},\Omega_{y_j}) =
\sum_{\ell_1\ell_2}
{\cal R}^{KL}_{\ell_x'\ell_y'\rightarrow \ell_1\ell_2}(\varphi_k)
\nonumber \\ \times
 N_{n\ell_1\ell_2} \sin^{\ell_1}\alpha_i \cos^{\ell_2}\alpha_i
P^{\ell_1 + 1/2, \ell_2 + 1/2}_n(\cos(2\alpha))
Y_{\ell_{1} \ell_{2}}^{LM_L}(\Omega_{x_i},\Omega_{y_i}) \; ,
\end{eqnarray}

\begin{equation}\label{a22}
 N_{n\ell_x\ell_y} =  \left( \frac{n! (n+\ell_x+\ell_y+1)! 
 \; 2(2n+\ell_x+\ell_y + 2)}
 {\Gamma(n+\ell_x+3/2)\Gamma(n+\ell_y+3/2)}\right)^{1/2}
\end{equation}

\begin{equation}\label{a23}
     Y_{\ell \ell}^{00}(\Omega_{x_i},\Omega_{y_i}) 
   Y_{\ell_{1} \ell_{2}}^{LM}(\Omega_{x_i},\Omega_{y_i}) =
  \sum_{\ell_x\ell_y}{\cal B}_{\ell_x\ell_yL}^{\ell_1\ell_2\ell}
   Y_{\ell_{x} \ell_{y}}^{LM}(\Omega_{x_i},\Omega_{y_i})  \;,
\end{equation}

\begin{eqnarray} \label{a25}
   {\cal B}_{\ell_x\ell_yL}^{\ell_1\ell_2\ell} =
    (-1)^{L+\ell_1+\ell_y+\ell}
   {\sqrt{(2\ell+1)(2\ell_1+1)(2\ell_2+1)}\over 4\pi}
 \nonumber \\ \times   
 \langle \ell_10\ell0|\ell_x0 \rangle  \langle \ell_20\ell0|\ell_y0 \rangle
   \left\{\begin{array}{lll}
      \ell_x&\ell_1&\ell\\
      \ell_2&\ell_y&L
   \end{array}\right\} \;,
\end{eqnarray}
where $Y_{\ell_{x} \ell_{y}}^{LM_L}$ is defined in eq.(\ref{e61}),
$N_{n\ell_1\ell_2}$ are normalzation constants for the Jacobi
polynomials $P^{\ell_1 + 1/2, \ell_2 + 1/2}_n(\cos(2\alpha))$, $\{\}$
and $\langle \rangle$ are the 6J-symbols and the Clebsch-Gordon
coefficients and the coefficients ${\cal
R}^{KL}_{\ell_x'\ell_y'\rightarrow \ell_1\ell_2}(\varphi_k)$ are the
so-called Raynal-Revai coefficients \cite{ray70} with
$K=\ell_x'+\ell_y'=2n+\ell_1+\ell_2$.

Then by combining eqs.(A\ref{a7}), (A\ref{a19})-(A\ref{a25}) we obtain
\begin{eqnarray} \label{a21}
\frac{\phi_{n \ell_x^\prime \ell_y^\prime L s_x^\prime S}^{(j) }
(\alpha_j)}{\sin(2\alpha_j)} 
Y_{\ell_{x'} \ell_{y'}}^{LM_L}(\Omega_{x_j},\Omega_{y_j}) =
\sum_{\ell}  A_{\ell}^{\ell_x'\ell_y'L}(\alpha_i) \frac{4\pi}{2\ell + 1}
Y_{\ell \ell}^{00}(\Omega_{x_i},\Omega_{y_i})
\nonumber \\ \times
\sum_{\ell_1\ell_2} \frac{ N_{n\ell_1\ell_2}}{N_{0\ell_x'\ell_y'}}
{\cal R}^{KL}_{\ell_x'\ell_y'\rightarrow \ell_1\ell_2}(\varphi_k)
 \sin^{\ell_1}\alpha_i \cos^{\ell_2}\alpha_i
P^{\ell_1 + 1/2 \ell_2 + 1/2}_n(\cos(2\alpha_i))
\nonumber \\ \times
Y_{\ell_{1} \ell_{2}}^{LM_L}(\Omega_{x_i},\Omega_{y_i}) =
\sum_{\ell_x\ell_y}{\cal C}_{\ell_x,\ell_y}^{\ell_x'\ell_y'L}(\alpha_i)
   Y_{\ell_x\ell_y}^{LM}(\Omega_{x_i},\Omega_{y_i}) \;,
\end{eqnarray}
where the expansion coefficients ${\cal C}$ are given by
\begin{eqnarray}\label{a27}
   {\cal C}_{\ell_x \ell_y}^{\ell_x'\ell_y'L}(\alpha_i)
   =\sum_\ell A_\ell^{\ell_x'\ell_y'L}(\alpha_i){4\pi\over 2\ell+1}
 \sum_{\ell_1\ell_2}
 \frac{N_{n\ell_1\ell_2}}{N_{0\ell_x'\ell_y'}}
\nonumber \\ \times
{\cal R}^{KL}_{\ell_x'\ell_y'\rightarrow \ell_1\ell_2}(\varphi_k)
 \sin^{\ell_1}\alpha_i \cos^{\ell_2}\alpha_i
P^{\ell_1 + 1/2, \ell_2 + 1/2}_n(\cos(2\alpha_i))
   {\cal B}_{\ell_x\ell_yL}^{\ell_1\ell_2\ell} \;.
\end{eqnarray}

Finally we have therefore the desired expression for eq.(\ref{e75})
\begin{eqnarray}\label{a31}
R_{ij}^{\ell_x \ell_y \ell_x^\prime \ell_y^\prime L}
\left[\frac{\phi_{n \ell_x^\prime \ell_y^\prime L s_x^\prime S}^{(j) }(\rho,
\alpha_j)}{\sin(2\alpha_j)}\right] =
\int {\rm d}\Omega_{x_i} {\rm d}\Omega_{y_i} 
\left[Y_{\ell_x \ell_y}^{LM_L}(\Omega_{x_i},\Omega_{y_i})\right]^*
 \nonumber \\
\times \frac{\phi_{n \ell_x^\prime \ell_y^\prime L s_x^\prime S}^{(j) }(\rho,
\alpha_j)}{\sin(2\alpha_j)}
Y_{\ell_{x}^\prime \ell_{y}^\prime}^{LM_L}(\Omega_{x_j},\Omega_{y_j}) =
{\cal C}_{\ell_x \ell_y}^{\ell_x'\ell_y'L}(\alpha_i) \; .
\end{eqnarray}
This completes the general derivation of the expression for the
transformation of the wave function from one set of Jacobi coordinates
to another.

The large-distance expansion is now found by first approximating
eq.(A\ref{a3}) for large $\rho$ and small $\alpha_i$ as $\alpha_j=
\varphi_k + \alpha_i \cos\gamma$. Then by using $\alpha_i << 1$ in
eq.(A\ref{a13}) we obtain
\begin{equation}\label{a15} 
A_{\ell}^{\ell_x'\ell_y'L}(\alpha_i=0)  =
{\phi_{n \ell_x^\prime \ell_y^\prime L s_x^\prime S}^{(j)} (\varphi_k) 
 \over \sin(2\varphi_k) \sin^{\ell_x'}\varphi_k \cos^{\ell_y'}\varphi_k}
 \delta_{\ell0} \; ,
\end{equation} 
which through eqs.(A\ref{a27}) and (A\ref{a25}) implies that
\begin{eqnarray}\label{a33} 
{\cal C}_{\ell_x \ell_y}^{\ell_x'\ell_y'L}(\alpha_i=0)&=&4 \pi \delta_{\ell_x0}
A_{0}^{\ell_x'\ell_y'L}(\alpha_i=0) 
 \frac{N_{n 0 L}}{N_{0\ell_x'\ell_y'}}
P^{1/2, L + 1/2}_n(1)
{\cal R}^{KL}_{\ell_x'\ell_y'\rightarrow 0 L}(\varphi_k)
 {\cal B}_{0LL}^{0L0} 
 \nonumber \\ 
&=& \frac{N_{n 0 L}}{N_{0\ell_x'\ell_y'}} P^{1/2, L + 1/2}_n(1)
 \frac{\phi_{n \ell_x^\prime \ell_y^\prime L s_x^\prime S}^{(j)} 
(\varphi_k)}{\sin(2\varphi_k)}
\frac{{\cal R}^{KL}_{\ell_x'\ell_y'\rightarrow 0 L}(\varphi_k)}
{\sin^{\ell_x'}\varphi_k \cos^{\ell_y'}\varphi_k} \; .
\end{eqnarray} 
For $\ell_x'=0, \ell_y'=L, n=0$ we can simplify the expression by
using $P^{1/2, L + 1/2}_0(1)=1$, ${\cal R}_{0L \rightarrow
0L}^{LL}(\varphi_k)$ = $(-1)^L \cos^{L}(\varphi_k)$. We then obtain
eq.(\ref{e80}). The angular wave functions corresponding to non-zero
$\ell_x'$ rapidly approach zero for large $\rho$-values and the their
contributions are therefore here assumed to be zero.

Thus for short range interactions for large $\rho$ and therefore also
for small $\alpha$ only $l_{x_i}=0$ components receive contributions
from the rotated wave functions from the other Faddeev components.
All partial waves with $l_{x_i}>0$ can then be solved
independently. The remaining three components with $l_{x_i}=0,
i=1,2,3$ must be solved as a set of coupled equations.

\end{document}